\documentclass[conference]{IEEEtran}

\IEEEoverridecommandlockouts

\usepackage{amsmath,amssymb,amsfonts}
\usepackage{algorithmic}
\usepackage{graphicx}
\usepackage{textcomp}
\usepackage{xcolor}
\usepackage{hyperref}
\usepackage{url}
\usepackage{xspace}
\usepackage{booktabs}
\usepackage{siunitx}
\usepackage{multirow}
\usepackage{cleveref}
\usepackage[utf8]{inputenc}
\usepackage{listings}
\usepackage{url}
\usepackage{longtable}
\usepackage{xcolor}
\usepackage{colortbl} 
\usepackage{ragged2e}
\usepackage{makecell}
\usepackage{tcolorbox}
\usepackage{natbib} 
\usepackage{float}
\usepackage{utfsym}
\usetikzlibrary{calc}
\renewcommand{\arraystretch}{1.1} 
\usepackage{subcaption}
\usepackage[commandnameprefix=always,defaultcolor=black]{changes} 
\usepackage[scaled=1.0]{inconsolata}    
\usepackage{siunitx}

\definecolor{darkgreen}{rgb}{0.0,0.6,0.3}
\definecolor{darkred}{rgb}{0.8,0.2,0.1}
\definecolor{verybrightgray}{rgb}{0.98,0.98,0.98}
\definecolor{yellow}{rgb}{0.7,0.6,0.0}
\definecolor{pink}{rgb}{0.8,0.5,0.7}
\definecolor{cyan}{rgb}{0.1,0.7,0.7}
\definecolor{mauve}{rgb}{0.6,0.1,0.7}

\lstdefinelanguage{Solidity}{
  keywords={
    contract, function, public, private, internal, external, interface, type,
    view, pure, returns, return, if, else, true, false,
    for, while, do, break, continue, emit, event,  new, delete,
    override, virtual, immutable, memory, storage, calldata,
    require, revert, assert, import, using, library, is, enum,
    struct, constructor, modifier, assembly, error
  },
  morekeywords={[2]receive,fallback,call,delegatecall,payable,extcodesize},
  morekeywords={[3]msg, tx, block, abi, this, super},
  morekeywords={[4]mapping, address, uint, uint8, uint16, uint32, uint64, uint128, uint256, int, int8, int16, int32, int64, int128, int256,    bool, string, bytes, bytes1, bytes32,ERC20},
  morekeywords={[5]nonReentrant,isHuman,onlyOwner},
  keywordstyle=\color{darkred}\bfseries,
  keywordstyle={[2]\color{darkgreen}\bfseries},
  keywordstyle={[3]\color{mauve}\bfseries},
  keywordstyle={[4]\color{blue}\bfseries},
  keywordstyle={[5]\color{black}\bfseries},
  morestring=[b]",
  comment=[l]{//},
  morecomment=[s]{/*}{*/},
  commentstyle=\color{gray}\ttfamily\itshape,
  stringstyle=\color{pink},
  basicstyle=\ttfamily\footnotesize,
  breaklines=true,
  columns=flexible,
  showstringspaces=false,
  tabsize=2
}

\newcommand{\eg}{e.g.\@\xspace}
\newcommand{\ie}{i.e.\@\xspace}

\newcommand{\code}[1]{\lstinline|#1|}

\let\oldtexttt\texttt
\renewcommand{\texttt}[1]{{\small\oldtexttt{#1}}}

\title{Reentrancy Detection in the Age of LLMs}

\begin{document}

\author{

\IEEEauthorblockN{1\textsuperscript{st} Dalila Ressi}
\IEEEauthorblockA{\textit{Ca' Foscari University of Venice} \\
Venice, Italy \\
dalila.ressi@unive.it}
\and
\IEEEauthorblockN{2\textsuperscript{nd} Alvise Spanò}
\IEEEauthorblockA{\textit{Ca' Foscari University of Venice} \\
Venice, Italy \\
alvise.spano@unive.it}
\and
\IEEEauthorblockN{3\textsuperscript{rd} Matteo Rizzo}
\IEEEauthorblockA{\textit{Ca' Foscari University of Venice} \\
Venice, Italy \\
matteo.rizzo@unive.it}
\and
\IEEEauthorblockN{\hspace{3.5cm}4\textsuperscript{th} Lorenzo Benetollo}
\IEEEauthorblockA{\hspace{3.5cm}\textit{University of Camerino} \\
\hspace{3cm}Camerino, Italy \\
\hspace{3cm}lorenzo.benetollo@unicam.it}

\and
\IEEEauthorblockN{5\textsuperscript{th} Sabina Rossi}
\IEEEauthorblockA{\textit{Ca' Foscari University of Venice} \\
Venice, Italy \\
sabina.rossi@unive.it}

}

\maketitle

\begin{abstract}
Reentrancy remains one of the most critical classes of vulnerabilities in Ethereum smart contracts, yet widely used detection tools and datasets continue to reflect outdated patterns and obsolete Solidity versions. 
This paper adopts a dependability-oriented perspective on reentrancy detection in Solidity 0.8+, assessing how reliably state-of-the-art static analyzers and AI-based techniques operate on modern code by putting them to the test on two fronts.
We construct two manually verified benchmarks: an Aggregated Benchmark of 432 real-world contracts, consolidated and relabeled from prior datasets, and a Reentrancy Scenarios Dataset (RSD) of \chadded{143} handcrafted minimal working examples designed to isolate and stress-test individual reentrancy patterns. 
We then evaluate 12 formal-methods-based tools, 10 machine-learning models, and 9 large language models. 
On the Aggregated Benchmark, traditional tools and ML models achieve up to 0.87 F1, while the best LLMs reach 0.96 in a zero-shot setting.
On the RSD, most tools fail on multiple scenarios, the top performer achieving an F1 of 0.76, whereas the strongest model attains 0.82.
Overall, our results indicate that leading LLMs outperform the majority of existing detectors, highlighting concerning gaps in the robustness and maintainability of current reentrancy-analysis tools.
\end{abstract}


\section{Introduction}
\label{sec:intro}

Reentrancy is the archetypal Ethereum vulnerability. Since the DAO exploit in 2016 \cite{feichtinger2024sok}, where an attacker drained millions of dollars worth of Ether by recursively invoking a withdrawal function, it has remained a central concern for auditors, tool builders, and platform operators. Recent estimates attribute approximately \$908.6M in losses to reentrancy-based attacks \chadded{\cite{yang2024uncover,atzei2017survey,perez2021smart}}.
Over the years, dozens of detection tools, ranging from symbolic execution engines \chadded{\cite{luu2016making} in conjunction with SMT solvers \cite{kalra2018zeus}} to abstract interpreters and pattern-based analyzers \chadded{\cite{tsankov2018securify,feist2019slither}}, have been introduced to identify reentrancy flaws before deployment.
\chadded{Additionally, a number of tools perform runtime monitoring \cite{rodler2019sereum}, fuzzing \cite{nguyen2020sfuzz}, tracing \cite{zhang2020txspector} and other dynamic strategies \cite{ferreira2020aegis} to detect a variety of vulnerabilities after deployment, including reentrancy.}

However, the dependability of this tool ecosystem has degraded over time. Solidity has evolved substantially, especially in version 0.8.x series, introducing new language features, revised error semantics, and modernized contract architectures. Many detectors have not kept pace: some support only legacy compiler versions, others crash on contemporary syntax, and several implement outdated notions of reentrancy tied to the original ``checks-effects-interactions'' (CEI) formulation \cite{liu2025reentrancy}. As a result, the long-term reliability of these tools, once viewed as stable components in the smart contract security pipeline, is increasingly uncertain.

This deterioration also affects machine-learning-based detectors. Most ML models are trained on datasets labeled by these same static tools. When detectors disagree, fail to run, or produce inconsistent warnings, these errors propagate directly into the training labels. The result is an ecosystem where ML detectors inherit tool-specific noise, outdated assumptions, and false positives, while little is known about their behavior when evaluated on clean, manually verified data.

The goal of this paper is to quantify the robustness, consistency, and long-term reliability of existing tools and AI-based techniques.
\chadded{We distinguish between \emph{reliability} and \emph{robustness} of reentrancy detectors.
Reliability refers to the correctness of predictions, i.e., the ability to accurately detect vulnerabilities while minimizing false positives and false negatives. Robustness denotes the capability of a detector to remain functional across evolving Solidity versions, language features, and contract patterns. 
Together, these properties determine the overall \textit{dependability} of a detection approach.
}

Our evaluation relies on two complementary datasets.
First, we build an \emph{Aggregated Benchmark} by consolidating and manually verifying contracts from three prior academic datasets whose authors claim manual labeling. Although this benchmark includes both real contracts and injected vulnerabilities, every entry was revalidated through multi-reviewer inspection and LLM-assisted explanation analysis, yielding a substantially higher-confidence alternative to tool-labeled corpora commonly used in ML research.
Second, we construct the \emph{Reentrancy Scenarios Dataset} (RSD), a handcrafted suite of minimal contracts (less than 30 lines) covering up-to-date Solidity features and patterns, such as error handling, advanced modifier-based guards, event emissions and code refactorings that obscure syntactic CEI guidelines. 
RSD is intentionally not representative of real-world prevalence: its purpose is to stress-test detectors under semantic corner cases, evolving language constructs and new habits within the dev community.

\chadded{These datasets enable an unprecedented evaluation of the dependability of 12 existing detection tools in the modern context. }
\emph{Our findings reveal a fractured and unreliable ecosystem}, with many tools failing outright on modern contracts, disagreeing sharply with one another, or producing high rates of false positives and false negatives even on relatively simple patterns.
\chadded{In addition, we test 10 traditional ML models and 9 LLMs.}
Several LLMs exhibit strong zero-shot performance across both datasets, achieving high accuracy without version-specific maintenance or handcrafted rules. 
Their explanations additionally help identify mislabeled edge cases and clarify ambiguous scenarios.

To structure this investigation, we articulate four research questions:

\begin{itemize}
    \item \textbf{RQ1:} How reliable and robust are existing reentrancy detection tools when applied to modern Solidity codebases and evaluated under stress-testing scenarios, given their lack of maintenance and outdated compiler support?

    \item \textbf{RQ2:} How consistent are the warnings and vulnerability reports produced by different tools when applied to the same contracts? 

    \item \textbf{RQ3:} Do general-purpose LLMs demonstrate higher \chadded{dependability} and lower maintenance cost compared to static analysis and ML-based detectors? 

    \item \textbf{RQ4:} Can LLM-generated explanations assist human experts in identifying mislabeled or ambiguous cases, thus improving dataset quality? 
\end{itemize}

Overall, our results demonstrate that the reliability of conventional reentrancy detectors has significantly eroded as Solidity evolved, calling into question their suitability as ground truth for both developers and ML pipelines. At the same time, LLMs (despite lacking formal guarantees) offer surprising robustness and semantic generality, outperforming outdated static analyzers on both stress tests and manually verified benchmarks \cite{chen2025chatgpt,ding2025smartguard}. These findings underscore the need for new hybrid approaches that combine the principled guarantees of formal tools with the adaptability of modern AI systems, supported by high-quality human-verified datasets rather than legacy tool outputs.

\subsection{Contributions}
\label{sec:contributions}

\begin{itemize}
    \item \textbf{A dependability-driven assessment of reentrancy detectors.}
    We provide the first systematic evaluation of the long-term dependability of existing reentrancy detection tools, revealing widespread fragility caused by maintenance decay, outdated compiler support, and inconsistent behavior across detectors.

    \item \textbf{Two complementary datasets enabling reliability and stress-test evaluation.}
    We release (i) the \emph{Reentrancy Scenarios Dataset (RSD)}, a curated stress-testing suite covering modern and challenging Solidity constructs, and (ii) an \emph{Aggregated Benchmark} built from three manually verified academic datasets. Together, these datasets enable both realistic reliability assessment and controlled robustness evaluation.

    \item \textbf{LLMs as a robust alternative and as human-in-the-loop assistants.}
    We show that several large language models outperform traditional detectors in accuracy and robustness while requiring no maintenance. Furthermore, their explanations help human auditors identify mislabeled or ambiguous cases, improving dataset quality and annotation reliability.
\end{itemize}


All code, datasets, and prompts are available in an anonymized repository.\footnote{Anonymized repository: \url{https://anonymous.4open.science/r/reentrancy-detection-benchmarks-A1D7}}

\section{Threat Model}
\label{sec:threat-model}


We consider Ethereum smart contracts written in Solidity and deployed on an EVM-compatible blockchain.
Contracts are compiled to bytecode and interact via message calls that may optionally transfer Ether.
In Solidity, external calls can be expressed either through a low-level primitive:
\begin{lstlisting}[language=Solidity]
target.call{value: amount}(data)
\end{lstlisting}

where \lstinline{target} is an \lstinline{address}, \lstinline{amount} is an integral expressing the amount of Ether being transferred, and \lstinline{data} encodes both the callee function name and actual arguments; or via a high-level call:
\begin{lstlisting}[language=Solidity]
I(target).f(args..)
\end{lstlisting}

where \lstinline{I} is an interface type (\eg ERC20), \lstinline{f} is a function name and \lstinline{args..} is the argument list.
Both syntaxes are compiled into the same \texttt{CALL} bytecode instruction; thus, they are equivalent at the runtime level.
A transaction may trigger a chain of contract invocations, including callbacks into the original caller contract - this is what reentrancy is about.
The contract state consists of its persistent storage variables and its internal balance.
We assume that the underlying blockchain provides standard consensus and integrity guarantees and that EVM opcodes behave as specified.

\subsection{Attacker Capabilities and Goals}

The adversary controls one or more externally owned accounts and can deploy arbitrary malicious contracts. The attacker can:
\begin{itemize}
    \item invoke any public or external function of the victim contract in arbitrary order, with chosen calldata and Ether values;
    \item implement fallback or \texttt{receive} functions that re-enter the victim during an external call;
    \item create other contracts that can invoke any function of the victim in any order at their construction time.
\end{itemize}

The attacker \emph{cannot} break cryptographic primitives, bypass the consensus protocol, forge signatures, or alter historical blockchain state. We focus solely on vulnerabilities arising from the contract logic and its interaction patterns.

The adversary's goal is to exploit \emph{reentrancy}: gaining multiple, nested entries into the victim contract within the same transaction so as to violate intended invariants. Typical consequences include draining Ether or tokens, inflating balances, tricking a contract to pay more times than intended, or otherwise reaching a contract state that would be impossible under any non-reentrant execution schedule.

\subsection{Operational Definition of Reentrancy}

We consider a contract vulnerable to reentrancy if, during the execution of one of its entry points:

\begin{enumerate}
\item it performs an external call to attacker-controlled code (e.g., via \texttt{call}, high-level interface calls, or \texttt{delegatecall});
\item upon return from that external call, it performs a state update that depends on the attacker’s interaction (e.g., modifying storage associated with \texttt{msg.sender});
\item and an attacker can re-enter the contract during the external call and drive it to a final state that \emph{cannot} be reproduced by any sequence of the same calls executed without reentrancy.
\end{enumerate}

Conditions (1)-(2) represent the minimal execution structure that \emph{permits} reentrancy; condition (3) characterizes when such a structure is \emph{exploitable}, i.e., when reentrant execution leads to a semantically different final state. This definition generalizes the classical checks–effects–interactions (CEI) violation pattern: CEI-compliant code is safe by construction, but CEI alone is insufficient to cover subtle vulnerable patterns such as read-only reentrancy, proxy-mediated flows, or semantics that intentionally place a state update after an external call.
This definition also highlights a central challenge: detecting reentrancy requires semantic reasoning, yet most existing datasets and tools rely on brittle syntactic rules.





\section{Motivation}
\label{sec:motivation}

A central challenge in evaluating reentrancy detectors is that no agreed-upon definition of reentrancy exists in practice. During our preliminary investigation, we found that both academic papers and widely used tools implement incompatible, outdated, and often underspecified heuristics. As a result, the ecosystem lacks a reliable ground truth: detectors disagree systematically, ML datasets inherit contradictory labels \cite{ressi2024vulnerability}, and many entries in “manually verified” datasets are not manually verified at all \cite{cgt}.

\medskip
\paragraph{Why manually labeled datasets were our starting point}
To avoid circularity in our evaluation (where tools are assessed on data labeled by other tools) we began with the only three datasets that explicitly claim human validation: the Consolidated Ground Truth dataset (CGT)~\cite{cgt}, the HuangGai dataset~\cite{huanggai}, and the large-scale reentrancy dataset used in~\cite{reentrancy_study}.
However, the CGT paper itself documents inconsistencies in prior manual labeling efforts, showing that the same contract is often labeled differently across publications.
This forced us to reexamine each dataset, rather than rely on legacy annotations implicitly shaped by outdated detectors.

\medskip
\paragraph{Why we extracted the rules of existing tools}
To understand the sources of disagreement, we analyzed how the main reentrancy analyzers actually define the vulnerability.
This analysis revealed a fragmented landscape: every tool encodes its own notion of reentrancy, often tied to pre-0.8 compiler behaviors or the 2016 DAO exploit. We selected a group of vulnerability detection tools from the CGT study and integrated it with other peer-reviewed tools as well as tools often used to label smart contracts for the training of AI models.

We briefly summarize the core rule each tool uses, demonstrating how inconsistent the definitions are.
Symbolic-execution tools such as \textbf{Mythril}~\cite{githubGitHubConsenSysDiligencemythril} and \textbf{Conkas}~\cite{githubGitHubSmartbugsconkas} flag reentrancy when a low-level \texttt{CALL} forwards more than 2,300 gas and is followed by an \texttt{SSTORE}, though Mythril ignores constructor calls and precompiled addresses, whereas Conkas checks SMT-feasible storage dependencies.
\textbf{Oyente} and Oyente+~\cite{luu2016making} treat a call as reentrant when its path condition remains satisfiable after hypothetical recursive execution.
Declarative or pattern-based analyzers diverge even more.
\textbf{CCC}~\cite{weiss2024analyzing} detects reentrancy via Code Property Graph queries requiring a user-influenced path from an external call to a subsequent state update.
\textbf{Securify}~\cite{tsankov2018securify} issues violation patterns whenever an \texttt{SSTORE} depends on a preceding \texttt{CALL}.
\textbf{Slither}~\cite{feist2019slither} expands the CEI rule into five distinct variants: counting event emissions, unlimited-gas calls, or even certain benign effects as reentrancy.
\textbf{Solhint} \cite{solhintGithub} 
 reduces reentrancy to pure CEI violations: any state write after an external call triggers a warning.
\textbf{Vandal}~\cite{brent2018vandal} uses Datalog queries on decompiled bytecode and flags a \texttt{CALL} as reentrant when it forwards enough gas, lacks a mutex, and is followed by state updates.
Dynamic tools implement yet another definition.
\textbf{Confuzzius}~\cite{torres2021confuzzius} flags reentrancy only when a concrete execution trace includes a \texttt{CALL} with  more than 2,300 gas, an \texttt{SLOAD} before it, and an \texttt{SSTORE} to the same slot after it—an explicit encoding of the DAO exploit.
\textbf{sFuzz}~\cite{nguyen2020sfuzz} deploys contracts to a sandbox chain and labels a contract vulnerable only if a synthesized attacker contract successfully reenters during fuzzing. 

For some of these tools, the precise rules were not properly documented, and we had to look at the implementation to infer the exact mechanism applied. In other cases, the paper reported the detection of reentrancy, but the rule was not implemented in the provided tool \cite{tikhomirov2018smartcheck}. The complete list of the tools we examined can be seen in \Cref{tab:tools}.

\newcommand{\Source}{\chadded{Source}}
\newcommand{\Bytecode}{\chadded{Bytecode}}
\newcommand{\Runtime}{\chadded{Runtime}}
\newcommand{\Mixed}{\chadded{Mixed}}

\newcommand{\crossmark}{\scalebox{0.8}{\usym{2613}}}
\newcommand{\working}{\checkmark}
\newcommand{\notworking}{\crossmark}
\newcommand{\NA}{N/A}

\begin{table*}[ht]
\centering
\caption{Comparison of analyzers for reentrancy detection.}
\label{tab:tools}
\resizebox{\textwidth}{!}{
\rowcolors{1}{white}{gray!20}

\begin{tabular}{|l|l|c|c|c|r|r|}
\hline
\textbf{Tool} & \textbf{Version/Commit} & \textbf{Working} & \textbf{\chadded{Input}} & \textbf{Techniques} & \textbf{Paper} & \textbf{Repo}
\\
\hline

Aderyn 
& v0.6.5 & \working & \Source
& Abstract syntax tree traversal and rule-based static pattern matching
& \NA & \cite{aderynGithub}
\\

AutoAR 
& \NA & \NA & \Mixed
& Program dependency graph construction and inter-procedural static analysis
& \cite{autoar2025} & -
\\

CCC 
& \#c531ae3 & \working & \Bytecode
& Code property graph analysis with Cypher query patterns
& \cite{weiss2024analyzing} & \cite{cccGithub}
\\

ConFuzzius
& \#4315fb7 v0.0.1 & \working & \Runtime
& Hybrid approach combining symbolic execution, syntax inspection, SMT solving, and fuzzing
& \cite{torres2021confuzzius} & \cite{confuzziusGithub} 
\\

Conkas
& \#4e0f256 & \working & \Runtime
& Symbolic execution with syntactic and SMT-based constraint analysis
& \cite{veloso2023conkas} & \cite{smartbugsConkasGithub}
\\

ContractWard
& \NA & \working & \Bytecode
& Machine-learning-based vulnerability classification with AST feature extraction
& \cite{wang2020contractward} & -
\\

eThor 
& 2023 & \notworking & \Bytecode
& Abstract interpretation, Horn clause resolution, and SMT solving
& \cite{schneidewind2020ethor} & -
\\

DefectChecker
& \NA & \working & \Bytecode
& Abstract syntax tree analysis with supervised machine learning
& \cite{chen2022defectchecker} & -
\\

Manticore 
& v0.3.7 & \notworking & \Bytecode
& Symbolic execution and SMT-based path exploration
& \cite{mossberg2019manticore} & \cite{manticoreGithub}
\\

Mythril 
& \chadded{v0.24.8} & \working & \Bytecode
& Symbolic execution with constraint solving and pattern-based vulnerability detection
& - & \cite{mythrilGithub}
\\

NPChecker
& \NA & \NA & \Bytecode
& Inter-procedural static analysis with constraint solving
& \cite{wang2019npchecker} & -
\\

Oyente+ 
& \#060ca34 & \working & \Bytecode
& Symbolic execution and SMT constraint analysis
& \cite{luu2016oyente} & \cite{oyentePlusGithub}
\\

Sailfish
& v0.1 & \working & \Bytecode
& Symbolic execution combined with static dependency and control-flow analysis
& \cite{bose2021sailfish} & -
\\

Securify 
& v1.0 & \working & \Bytecode
& Datalog-based data-flow and call-dependency analysis
& \cite{tsankov2018securify} & \cite{securifyGithub}
\\

Securify2 
& \NA & \notworking & \Bytecode
& Declarative intermediate-representation analysis using Datalog compliance and violation rules
& - & \cite{securify2Github}
\\

Sereum
& v1.0 & \NA & \Runtime
& Dynamic taint tracking on Ethereum bytecode during runtime execution
& \cite{rodler2019sereum} & -
\\

sFuzz 
& \#48934c0 & \working & \Runtime
& Feedback-guided fuzzing with transaction sequence generation
& \cite{nguyen2020sfuzz} & \cite{sfuzzGithub}
\\

Slither 
& v0.11.3 & \working & \Source
& Static pattern matching with inter-procedural data-flow and control-flow analysis
& \cite{feist2019slither} & \cite{slitherGithub}
\\

Smartcheck
& \NA & \notworking & \Source
& Abstract syntax tree pattern matching and heuristic regular expressions
& \cite{tikhomirov2018smartcheck} & -
\\

Solhint 
& v6.0.0 & \working & \Bytecode
& Linting rules and static pattern matching on Solidity source code
& - & \cite{solhintGithub}
\\

teEther
& \#04adf56 & \notworking & \Bytecode
& Symbolic execution and SMT-based exploit generation
& \cite{krupp2018teether} & \cite{teetherGithub}
\\

TotalSol
& \#04adf56 & \NA & \Mixed
& Multi-layer analysis combining control-flow, data-flow and inter-procedural dependency
& \cite{mishra2025totalsol} & -
\\

Vandal 
& \#d2b0043 & \working & \Bytecode
& Bytecode decompilation and Datalog-based static analysis
& \cite{brent2018vandal} & \cite{vandalGithub}
\\

\hline
\end{tabular}}

\end{table*}

\medskip
\paragraph{Why this fragmentation matters}
These rules are not minor variations, they are mutually incompatible interpretations of the same vulnerability.
Tools disagree on whether:
\begin{itemize}
    \item unlimited-gas high-level calls are equivalent to \texttt{call()}~\cite{luu2016making};
    \item \texttt{delegatecall} should always be treated as reentrant~\cite{schneidewind2020ethor};
    \item read-only reentrancy is a type of reentrancy vulnerability~\cite{zhang2024smartreco};
    \item mutexes or modifiers constitute adequate guards~\cite{feist2019slither}.
\end{itemize}

Many rules further depend on compiler behaviors that no longer hold in Solidity~0.8+, such as automatic underflow checks, fallback semantics, and storage access patterns. This explains why detectors often break on modern code.

\medskip
\paragraph{Implications for ground truth}
Because each dataset in the literature was labeled according to a specific tool’s rule (often undocumented\chadded{, especially in papers of tools detecting multiple vulnerabilities}) the resulting labels are inconsistent \chadded{\cite{cgt,ressi2024vulnerability,DurieuxEtAl2020ICSE}}.
This is why prior work frequently disagrees on whether a contract is vulnerable. Using such datasets for evaluation or ML training introduces tool-induced bias and artifacts \chadded{\cite{ressi2024vulnerability}}.

This fragmentation motivated our approach:
(1) extract and analyze the actual rules that tools encode;
(2) design a unified three-step labeling procedure grounded in the execution model;
(3) re-verify all contracts in existing manually labeled datasets;
(4) create a second dataset (RSD) to systematically exercise scenarios that existing detector rules mishandle.

\medskip
Both datasets, as described next, are thus explicitly aligned with our threat model and eliminate the tool-defined inconsistencies that have undermined prior evaluations.

\section{Dataset Construction}
\label{sec:datasets}

Because existing datasets often embed tool-induced inconsistencies \chadded{\cite{cgt,ressi2024vulnerability,DurieuxEtAl2020ICSE}}, we construct two new datasets aligned with the threat model above. One captures semantic stress tests, the other provides high-confidence labels for real-world contracts. Below, we describe their construction and the unified labeling procedure that ensures consistency across both.

\subsection{Aggregated Benchmark}
\label{sec:aggregated-benchmark}

To evaluate detectors on contracts that developers are likely to encounter in practice, we first assemble a high-confidence benchmark from the largest three prior datasets that explicitly claim manual validation: the Consolidated Ground Truth (CGT)~\cite{cgt},  HuangGai~\cite{huanggai}, and the reentrancy study by~\cite{reentrancy_study}. These sources were selected because they contain human-labeled reentrancy cases rather than tool-generated labels.
\chadded{In particular, CGT is itself an aggregated dataset, comprising 13 different datasets. Among these, some of the most well-known are ZEUS \cite{kalra2018zeus}, SolidiFI \cite{ghaleb2020effective}, SmartBugs Curated \cite{DurieuxEtAl2020ICSE} and SWC registry}
\footnote{\chadded{\url{https://swcregistry.io/}}}.

We aggregate all contracts from these datasets, deduplicate them, and filter out those that fail to compile under standard Solidity compiler versions (most of the contracts are written in Solidity v0.4.x or v0.5.x, with only a few in v0.6.x).
\chadded{Some contracts do not compile due to a missing pragma statement in the preamble; others do so due to random errors in the code, despite the exact compiler version being adopted.
}

This yields a pool of \num{73579} unique, compilable contracts, consisting of 145 potentially reentrant and \num{73434} potentially non-reentrant instances, summarized in \Cref{tab:dataset}.
As documented in~\cite{cgt}, even “manually labeled” corpora often contain contradictory or inconsistent annotations, motivating the need for re-verification under a consistent threat model. 

\medskip
\textbf{Manual relabeling under our operational definition.}
All contracts were manually inspected independently by three authors in multiple rounds, which eventually led to the definition of the procedure in \Cref{sec:procedure}. This procedure operationalizes reentrancy through a combination of external-call detection, post-call side effects, and semantic state divergence.
A final round was then performed to ensure adherence to the converged procedure. 
\chadded{Authors brought complementary expertise: one in programming languages and formal methods for reentrancy detection, one in smart contract development for Ethereum, and one in vulnerability detection and AI-based contract analysis. Disagreements stemmed from occasional misapplications of our reentrancy-labeling procedure (cf. \Cref{sec:procedure}), and were resolved through joint inspection of the more subtle cases.
}

\begin{table}[hb]
    \centering
    \caption{Aggregated Benchmark Dataset Composition.}
    \sisetup{
        group-separator={,},
        group-minimum-digits={4},
        detect-weight=true,
        detect-inline-weight=math
    }
    \label{tab:dataset}
    
    \setlength{\tabcolsep}{3pt}
    \resizebox{\columnwidth}{!}{
    {\footnotesize
    \begin{tabular}{@{} l r r r @{}}
        \toprule
        \addlinespace[0.5em]
        \multicolumn{1}{l}{\textbf{Contracts}} & 
        \multicolumn{1}{c}{\shortstack{\textbf{Initial}\\\textbf{Raw Count}}} & 
        \multicolumn{1}{c}{\shortstack{\textbf{Unique}\\\textbf{Contracts}}} & 
        \multicolumn{1}{c}{\shortstack{\textbf{Compilable}\\\textbf{Contracts}}} \\
        \midrule
        \quad \# Reentrant in CGT \cite{cgt} & 300 & 296 & 93 \\
        \quad \# Reentrant in HuangGai \cite{huanggai} & 22 & 22 & 21 \\
        \quad \# Reentrant in Reentrancy Study \cite{reentrancy_study} & 46 & 40 & 31 \\
        \midrule
        Total \textbf{Reentrant} (\textbf{before} manual verification) & 368 & 358 & 145 \\
        Total \textbf{Reentrant} (\textbf{after} manual verification) & {145} & 120 & \textbf{120} \\
        \midrule
        \quad \# Non-Reentrant in Reentrancy Study \cite{reentrancy_study} & \num{123169} & \num{118422} & \num{73434} \\
        
        \midrule
        Total \textbf{Non-Reentrant} (\textbf{before} manual verification) & \num{123169} & \num{118422} & \num{73434} \\
        Total \textbf{Non-Reentrant} (\textbf{after} manual verification) & {287} & 312 & \textbf{312} \\
        
        \midrule
        \textbf{Total in Aggregated Benchmark} & {} & 432 & \textbf{432} \\
        \bottomrule
    \end{tabular}
    }}
\end{table}

To enrich the dataset, we selected some contracts from the \num{73434} potentially non-reentrant contracts considered safe in~\cite{reentrancy_study}. The criteria used by the authors were that not all the tools they used to detect reentrancy on those contracts (Oyente, Mythril, Securify, Smartian \cite{choi2021smartian}, Sailfish) gave a negative response. 
We used their results to select contracts considered safe by at least 4 of those tools, then extracted a subset of this large batch with a sampling strategy that balances coverage with feasibility, while ensuring inclusion of patterns known to confuse analyzers (e.g., modifiers, inheritance, inline effects) by using regular expressions. 
\chadded{
The threshold of 4 agreeing tools was selected as a trade-off between label confidence and contract diversity. 
Requiring 5 tools produced an overly homogeneous subset dominated by low-level call patterns, while four preserved broader interaction types (including high-level calls) without significantly weakening consensus. 
A threshold of 3 tools was not adopted, as it increased the dataset size at the cost of reduced confidence in the ground-truth labels.}
After all rounds of manual inspection, numerous contracts originally labeled as reentrant were moved to the safe group, and vice versa. 

\medskip
\textbf{LLM-assisted verification.}
To further increase confidence, we queried several LLMs (gpt-4o, gpt-4.1, \texttt{o3-mini}, \texttt{o4-mini}) using a simple prompt requesting a label and explanation. Whenever \chadded{the output label of} an LLM disagreed with our current \chadded{label} annotation, the same three authors reviewed its explanation. In most cases, disagreements were hallucinations; in a minority, the LLM highlighted subtle behaviors we had overlooked. These cases were re-examined manually and relabeled if necessary. LLMs, therefore, did not label the dataset but served as an additional signal to identify edge cases.

\medskip
\textbf{Final benchmark.}
After this two-step verification (manual re-annotation plus LLM-assisted conflict review) the Aggregated Benchmark contains 432 high-confidence contracts: 122 reentrant and 314 safe. During the process, 28 of the 145 potentially reentrant contracts were relabeled safe, while 5 of the 291 sampled potentially safe contracts were relabeled reentrant. The resulting benchmark offers far higher label reliability than any of the original sources and reflects the operational definition established in our threat model.

\subsection{Reentrancy Scenarios Dataset (RSD)}
\label{sec:rsd}

Our analysis of the Aggregated Benchmark, combined with observations of tool behavior and LLM explanations, revealed recurring blind spots in existing datasets. These include limited coverage of modern Solidity 0.8.x features, insufficient representation of alternative external call mechanisms, and a near-complete absence of structurally diverse variants. To address these gaps, we constructed the Reentrancy Scenarios Dataset (RSD), a suite of \chadded{143} manually verified \chadded{Minimal Working Examples (MWEs)},

\begin{table*}[!ht]
    \centering
    \caption{Scenario taxonomy of the Reentrancy Scenarios Dataset (RSD). Categories group MWEs exhibiting similar structural patterns. }
    
    \resizebox{\textwidth}{!}{
    \rowcolors{1}{white}{lightgray!30}
    \setlength{\tabcolsep}{4pt} 
    \begin{tabular}{|c|l|c|c|c|l|p{3.4cm}|}
    \hline
        \textbf{Code} & \textbf{Name Prefix} & \textbf{\#Safe} & \textbf{\#Ree} & \textbf{Reentrancy Type }& \textbf{Description} & \textbf{Variants} \\
        \hline
        00 & Basic & \chadded{8} & \chadded{7} & single function 
        & Basic DAO in multiple variants and flavours
        & Emit, Error, Fold, Staticcall, Unchecked, Inline 
        \\ 
        00 & BasicConst & \chadded{1} & \chadded{1} & single function 
        & Basic DAO with constant address as target
        & 
        \\ 
        00 & BasicCross & \chadded{0} & \chadded{1} & cross function 
        & Basic DAO in a cross-function setting
        & 
        \\ 
        00 & BasicNoChecks & \chadded{1} & \chadded{1} & single function  
        & Basic DAO implementing CEI and omitting the checks part
        & 
        \\ 
        00 & BasicNoCall & \chadded{1} & \chadded{0} & single function  
        & Dummy contract without external calls
        & 
        \\ 
        01 & SingleMutex & 4 & 6 & single function 
        & DAO-like with flag-based mutex protection in a single function settings
        & Fold, Underflow 
        \\
        02 & CrossMutex & 4 & 4 & cross function 
        & DAO-like with flag-based mutex protection in a cross-function setting
        & Underflow, Unchecked 
        \\ 
        03 & SingleMod & 5 & 8 & single function
        & DAO-like with a modifier guard in a single function settings
        & Fold, Underflow 
        \\ 
        04 & CrossMod & 5 & 8 & cross function
        & DAO-like with a modifier guard in a cross-function setting
        & Fold, Underflow 
        \\ 
        05 & Send & 6 & 0 & single function 
        & DAO-like using send primitive (always safe)
        & Emit, Unchecked 
        \\
        06 & Transfer & 4 & 0 & single function 
        & DAO-like using transfer primitive (always safe)
        & Unchecked 
        \\ 
        07 & MixedSend & 3 & 3 & single function 
        & Send primitive followed by a low-level call 
        & Fold, Emit 
        \\ 
        08 & MixedTransfer & 2 & 2 & single function 
        & Transfer primitive followed by a low-level call 
        & Emit 
        \\ 
        09 & ERC20 & 7 & 10 & single function 
        & Donations via ERC20 tokens mixing various high-level calls
        & Mod, Inheritance, Staticcall 
        \\ 
        09 & ERC20OnlyOnce & 2 & 1 & single function
        & Donations via ERC20 tokens that can occur only once per user
        & ~
        \\ 
        09 & ERC20Staking & 3 & 3 & single function 
        & Staking application with ERC20 tokens 
        & 
        \\ 
        09 & ERC20StakingPull & 3 & 3 & single function
        & Staking application with ERC20 tokens including withdrawal
        & Mod 
        \\ 
        10 & OnlyOnce & 2 & 0 & single function
        & DAO-like where withdrawal can occur only once per user (always safe)
        & ~
        \\ 
        11 & Proxy & 3 & 0 & none
        & Proxy contract forwarding calls to a given target
        & Staticcall
        \\ 
        12 & OnlyOwner & 1 & 1 & cross function 
        & Special guard checking if the sender is the original owner of the contract
        & ~ 
        \\
        13 & Loop & 1 & 1 & single function 
        & Loop containing CEI-breaking code 
        & ~ 
        \\ 
        13 & LoopCrossMod & 3 & 4 & cross function
        & Loop containing CEI-breaking code in a cross-function setting
        & 
        \\ 
        13 & LoopCrossMutex & 3 & 4 & cross function
        & Loop guarded by mutexes
        & 
        \\ 
        14 & DelegateCall & 0 & 4 & delegatecall
        & Delegate calls are always vulnerable under our definition
        & ~ 
        \\
        15 & ReadOnly & 3 & 3 & cross contract
        & Modifications in the state of one contract make another vulnerable
        & Staticcall 
        \\ 
        \hline
    \end{tabular}}
    \label{tab:scenarios}

\end{table*}

The structure of RSD can be seen in \Cref{tab:scenarios}.
RSD is organized into scenario families (identified by the same ``code") reflecting well-known reentrancy patterns and real-world contracts.
Each category isolates a distinct source of potential reentrancy, such as breakages of the CEI pattern, misuse of guard mechanisms such as \texttt{nonReentrant} and \texttt{onlyOwner}, ERC20 token logic, proxy-based delegation, dangerous primitives such as \texttt{delegatecall} and so on.
For each scenario, we construct multiple variants by applying controlled syntactic and structural transformations or by simply injecting certain constructs.
Variants include function folding (\texttt{Fold}), error raising (\texttt{Error}), event emission (\texttt{Emit}), use of modifiers (\texttt{Mod}), contract inheritance (\texttt{Inherit}), inlining of side effects within a statement (\texttt{Inline}), use of \texttt{unchecked} blocks (\texttt{Unchecked}), use of low-level static call (\texttt{Staticcall}), and use of in-place subtraction operator (\texttt{Underflow}).
\chadded{These variants do not fabricate unrealistic behaviors; rather, they preserve the original reentrant control flow and EVM semantics while representing syntactic and structural variations commonly found in real-world contracts, precisely targeting patterns that challenge both static analyzers and LLMs.}

\Cref{tab:scenarios} summarizes the RSD by category, including the number of safe and vulnerable instances per scenario family, the reentrancy type involved, a short description, and the main variants explored.
Each category is identified by a code number and a name prefix.
We now showcase code examples only for the most representative scenarios, and refer the reader to the full dataset included in the supplementary material.

\medskip
\textbf{Basic (00).}
The simplest scenario features one function performing an external call that can reenter into itself\footnote{The state variable \texttt{bal} is defined once here and referenced from most snippets in this section.}.

\begin{lstlisting}{language=Solidity}
mapping (address => uint256) public bal;

function withdraw(uint256 amt) public {
  require(bal[msg.sender] >= amt, "No funds");
  (bool success, ) = msg.sender.call{value:amt}("");
  require(success, "Call failed");
  bal[msg.sender] -= amt;
}
\end{lstlisting}

The state update after the external call violates the CEI practice and leads to potential attacks.

\medskip
\textbf{SingleMod (03).}
To programmatically prevent reentrancy, contracts may implement a mutex-like mechanism using a global boolean flag. 
This flag is typically tested and set at the beginning of the code section that performs an external call, and reset in the epilogue.  
Unlike traditional mutexes in concurrent systems, atomicity is not required here, as the Ethereum execution model is inherently sequential.  
Using a mutex to guard reentrant functions enables developers to intentionally deviate from the CEI pattern, safely performing state updates \emph{after} external calls without exposing the contract to reentrancy vulnerabilities.
This pattern can be implemented in a couple of equivalent ways - either manually, or through a Solidity modifier that implements a reentrancy guard:

\begin{lstlisting}{language=Solidity}
modifier nonReentrant() {
  require(!flag, "Locked");
  flag = true;
  _;    // placeholder for the function body
  flag = false;
}
function withdraw(uint256 amt) nonReentrant public {
  require(bal[msg.sender] >= amt, "No funds");
  (bool success, ) = msg.sender.call{value:amt}("");
  require(success, "Call failed");
  bal[msg.sender] -= amt; // effect after call is ok
}
\end{lstlisting}

Although the snippet above illustrates a non-reentrant sample, flawed implementations of this pattern may still introduce reentrancy vulnerabilities, such as wrongly implementing the boolean flag management or forgetting the require statement.

\medskip
\textbf{CrossMod (04).}
This category includes contracts composed of multiple functions, where one performs an exploitable external call.  
Despite the reentrancy guard, an attacker can still re-enter other public functions:

\begin{lstlisting}{language=Solidity}
function transfer(address to, uint256 amt) public {
  require(bal[msg.sender] >= amt, "No funds");
  bal[to] += amt;   // reentering here alters state
  bal[msg.sender] -= amt;
}

function withdraw(uint256 amt) nonReentrant public {
  require(bal[msg.sender] >= amt, "No funds");
  (bool success, ) = msg.sender.call{value:amt}("");
  require(success, "Call failed");
  bal[msg.sender] -= amt;
}
\end{lstlisting}

Such scenarios may be exploited in subtler ways compared to direct single-function reentrancy: the reentrant manipulation of the contract state can result in unintended behaviors, such as altered payment amounts, even when the attack does not involve direct theft of funds.

\medskip
\textbf{Send/Transfer (05/06).}
Solidity provides the \code{send} and \code{transfer} primitives as methods of the \code{address} built-in type.
These must not be confused with other \code{send} or \code{transfer} methods belonging to interfaces (\eg ERC20) or third-party libraries.
The two primitives differ only for their signature: \code{send} returns a boolean indicating whether the operation has succeeded, while \code{transfer} simply reverts the transaction automatically when it fails.

\begin{lstlisting}{language=Solidity}
function withdraw(uint256 amt) public {
  require(bal[msg.sender] >= amt, "No funds");
  bool success = payable(msg.sender).send(amt);
  require(success, "Call failed");
  bal[msg.sender] -= amt; // effect after call is ok
}
\end{lstlisting}

Both primitives, however, do \emph{not} allow reentrancy despite invoking the receive/fallback function on the target address (if non-EOA), thanks to a hard limit in the gas at the avail of the target (2300 at the time of writing).
Such a small amount does not provide a malicious callee with enough gas to re-enter.

\medskip
\textbf{ERC20 (09).}
The ERC20 standard for managing fungible tokens on Ethereum is defined as a well-known interface in Solidity, which specifies a set of functions and events \cite{openzeppelinOpenZeppelinDocs}.
Contracts implementing this interface typically entail explicitly type-casting an \code{address} to the \code{ERC20} interface and then invoking its methods.

\begin{lstlisting}{language=Solidity}
mapping (address => bool) private donated;

function donate(address tok, address to,
                uint256 amt) public {
  require(!donated[msg.sender]);
  require(IERC20(tok).balanceOf(msg.sender) >= amt * 2, "Need twice");
  bool success = IERC20(tok).transfer(to, amt);       
  require(success, "Donation failed");
  donated[msg.sender] = true; // side-effect
}
\end{lstlisting}

Invoking \code{balanceOf} and \code{transfer} delegates execution to an external contract whose behavior is unknown, as it may re-enter the caller function or perform other malicious operations.
In general, scenarios based on ERC or any other interface can be reduced to the same case: an explicit type-cast from an \code{address} to a contract or interface type, when followed by a method invocation, is considered a \textbf{high-level call}.
And that is equivalent to a low-level call under the following reversible syntactic transformation:
$$
\begin{array}{c}
e_0.\text{\lstinline[basicstyle=\small]{call(abi.encodeWithSignature}}(``\mathtt{f}(\tau_1, .., \tau_n)", e_1, .., e_n))
\\
\rotatebox{90}{$\leftrightsquigarrow$}
\\
\mathtt{I}(e_0).\mathtt{f}(e_1, .., e_n)
\end{array}
$$
where expressions $e_0 : \mathtt{address}$, $e_i : \tau_i$ for all $i \in [1, n]$ and $\mathtt{I}$ is an interface type name defining the function signature $\mathtt{f}(\tau_1, .., \tau_n)$.
The two syntaxes make the Solidity compiler emit the same \code{CALL} instruction, hence they are equivalent from the point of view of the bytecode.
For analyzers processing Soldity code, though, they appear different.

\medskip
\textbf{OnlyOwner (12).}
Another commonly used reentrancy guard is the \code{onlyOwner} modifier, which restricts access to specific functions by ensuring that the caller is the contract's deployer (also known as the owner). 
\begin{lstlisting}{language=Solidity}
modifier onlyOwner() {
  require(msg.sender == owner, "Not authorized");
  _;
}
\end{lstlisting}

The modifier is often used in place of \code{nonReentrant}, although it just enforces an access control policy.
To truly avoid reentrant attacks, especially in cross-function scenarios, a mutex shared among all public/external functions is required.

\subsection{Labeling Procedure}
\label{sec:procedure}

\newcommand{\block}{\mathcal{S}}

Both the Aggregated Benchmark and the RSD have been labeled according to a unified three-step procedure, not intended as a full formalization of reentrancy but as a coherent operational definition ensuring consistency across datasets.
The first two steps rely on transformations and detection patterns that are syntactic in nature, aided by a slight amount of type information. 
The third step introduces the notion of contract state and ventures into the territory of semantics, refining the detection and ruling out some false positives.

Intuitively, given a Solidity contract, we begin by unfolding local functions (Step 1).
We then search in the unfolded code for external calls followed by side-effects (Step 2).
If none are found, the contract is classified as safe; otherwise, a \emph{candidate} reentrancy is identified and subjected to an additional semantic analysis (Step 3). 
This final step compares the contract state produced by executing the nested reentrant calls with the state obtained if the same calls were executed as independent transactions in sequence.
In other words, we contrast the outcome of mutually recursive reentrant executions with the outcome of their flattened, sequential counterparts.
If the two outcomes coincide, the contract is deemed safe (w.r.t. reentrancy); otherwise, it is vulnerable to reentrancy.

\chadded{
Compared to prior systems, our procedure captures both traditional CEI-breaking patterns and advanced cases (e.g., read-only and cross-contract reentrancy) under a unified criterion.
It diverges from traditional approaches \cite{luu2016making,atzei2017survey,perez2021smart} by systematically flagging delegatecalls as exploitable, while excluding emit and error from effects.
Also, a flat normalization of modifiers, local calls, and inheritance reduces constructs to a small core, avoiding ad hoc rules and limiting subtle false negatives.
}

\medskip
\textbf{Step 1. Unfolding.}
The first step performs a semantic-preserving transformation of a Solidity contract through the unfolding of local function calls and of a few other language constructs.
Local function calls exhibit no object expression on the left-hand side of the dot and have syntactic form $\mathtt{f}(e_1, ..., e_n)$ where $n \geq 0$, $e_1~ .. ~e_n$ are arguments of any type, and $\mathtt{f}$ is a function name identifier.
Notable exceptions are calls of the form $\mathtt{this.f}(e_1, ..., e_n)$, which the Solidity compiler treats as external calls and are therefore not unfoldable.

Function unfolding is a well-studied program transformation technique \cite{burstall1977transformation,turchin1986concept,partsch1983program} whereby a function call is systematically replaced with the body of its definition, with actual parameters substituted for formal ones.
Given a function $f(x_1 : \tau_1, .., x_n : \tau_n)~ \block$, with $\block = \{ S_1; ...; S_m \}$ being the function body, where $n \geq 0$, $m \geq 1$, $x_1 ~..~ x_n$ are parameters annotated with their types $\tau_1~ ..~ \tau_n$ and $S_1 .. S_m$ are statements, by unfolding each call expression in $\block$ we obtain a function $f'(x_1 : \tau_1, .., x_n : \tau_n)~ \block'$ with $\block' = \{ S'_1; ...; S'_k \}$, with $k \geq m$.
This process is repeated for each \emph{entry-point} in the input contract, \ie functions marked with \code{external} or \code{public}, as well as constructors, until the whole input contract consists of unfolded functions.

Unfolding also applies to the following language constructs, requiring just a trivial substitution mechanism:
\begin{itemize}
    \item \emph{Inheritance}: each member (state variable or function) defined in a superclass (parent contract) is embedded in the subclass (derived contract), and unfolding of local functions is applied to the resulting entry-points. The embedding is recursive and terminates at the root of the hierarchy.
    \item \emph{Modifiers} are unfolded by substituting the prologue and the epilogue code segments into the functions utilizing them and eventually applying local function unfolding to the result.
\end{itemize}

After unfolding, functions marked with \code{private} or \code{internal} are removed and only entry-points are left.

\medskip
\textbf{Step 2. Detecting Vulnerable External Calls.}
We define a mechanism for detecting violations of the CEI guideline in unfolded code.
The mechanism is based on both syntactic patterns in conjunction with type information; in particular, the types of Solidity expressions are required.

We introduce a limited set of \emph{blacklisted} primitives and language constructs performing external calls:

\begin{itemize}
\item \emph{Low-level calls}: expressions of form $e_0.\mathtt{call}(e_1)$ where expression $e_0 : \mathtt{address}$ and argument $e_1 : \mathtt{bytes}$.
\item \emph{High-level calls}: expressions of form $e_0.\mathtt{f}(e_1, ~..., ~e_n)$ where $n \geq 0$, $e_1~ .. ~e_n$ are arguments of any type, $\mathtt{f}$ is a function name identifier, $e_0: \tau$ is the result of an explicit type-cast (either in-place or earlier in the scope) of form $\tau(e)$, where $e : \mathtt{address}$.
\end{itemize}

The first bullet point detects proper invocations to the \texttt{call} primitive.
Types are required to rule out invocations of user-defined methods named \texttt{call}.
The second bullet point detects external calls that exhibit an address type-cast into an interface or contract type name.

For each unfolded entry-point function $f'(x_1 : \tau_1, .., x_n : \tau_n)~ \block'$ with $\block' = \{ S'_1; ...; S'_k \}$, $f'$ is considered \textbf{reentrant} if and only if \emph{both} the following conditions hold:
\begin{itemize}
    \item $\exists S'_i \in \block'$, for some $i \in [1,k]$, performing an \emph{external call} belonging to the set of blacklisted calls;
    
    \item \textbf{and} $\exists S'_j \in \block'$ such that $j > i$ performing some \emph{side effect} on a state variable.
\end{itemize}

The following statements are considered \textbf{side effects}:
\begin{itemize}
\item the assignment statement $e_1 = e_2$, where $e_2$ is an expression of any form and $e_1$ belongs to the following forms:
\begin{itemize}
    \item plain identifiers of form $x$ that belong to the contract state variables (local variables are ignored);
    \item field select $e.x$, where expression $e : \tau$ such that $\tau$ is a contract or struct type and $x$ is a field of $\tau$;
    \item array subscript $e[e']$, where $e : \tau[]$ and $e' : \mathtt{uint}$;
\end{itemize}

\item statements including a subexpression of form $e_0.\mathtt{delegatecall}(e_1)$ where expression $e_0 : \mathtt{address}$ and argument $e_1 : \mathtt{bytes}$.
\end{itemize}

Event emission (\code{emit}) and other special statements such as \code{require}, \code{assert}, \code{error}, and \code{revert} are \emph{not} considered side effects.
When Step 2 detects a side effect after an external call, the contract is marked as \emph{candidate} reentrant and Step 3 is necessary to rule out false positives; otherwise it is marked as \emph{safe} and the procedure stops.

\medskip
\textbf{Step 3. Contract State Comparison.}

We introduce the notion of \emph{state of a contract} as the tuple of state variables, \ie, global fields therein defined, plus the state of the contract internal balance.
Such state can either be valid or invalid due to an error ($\bot$).
\chadded{By error, we mean a transaction revert, whether deliberately triggered by the user code or induced by the EVM during execution.}

Let $\sigma_0$ denote the contract state prior to the invocation of a candidate reentrant function $f$, and let $\sigma_n$ denote the state resulting from $n$ reentrant invocations of $f$ within a single transaction.
Similarly, let $\sigma'_n$ denote the state produced by $n$ independent non-reentrant calls to $f$, each executed in separate transactions. 
A contract is vulnerable to reentrancy if and only if $\sigma_n \neq \sigma'_n$, meaning that chaining through reentrant calls induces an observable divergence in the state that would not arise under isolated executions.
Error states make no exception: when both executions terminate in error, \ie when $\sigma_n = \sigma'_n = \bot$, the contract is deemed safe.
Conversely, if either $\sigma_n$ or $\sigma'_n$ becomes unreachable due to an intermediate error, \ie a state transition $\sigma_i = \bot$ or $\sigma'_i = \bot$ for some $i < n$, then the contract is classified as reentrant.
\chadded{To be considered safe, a contract must therefore reach the same state, whether valid or not, in the same number of steps in both forms, \ie as a single transaction performing reentrant calls and as separate sequential transactions.}

\medskip
\chadded{While the procedure outlined above is presented at a semi-formal level, it is designed as a reproducible workflow for manual labelling by expert human analysts. For sufficiently simple programs with a manageable number of external calls, each step is practical to perform, including step 3, despite its semantic nature.
A complete example of application of the procedure is available as supplementary material}\footnote{\chadded{Relative path from repository root: \url{labeling\_precedure\_example.pdf}}}.

\section{Methodology}
\label{sec:methodology}

In this section we delve into the details of how we run the tool suite presented in \Cref{tab:tools} on our two datasets, and how we eventually extracted the relevant information from their output logs in order to compute the proper metrics.

\subsection{Experimental Setup}
\label{sec:experimental_setup} 

Experiments utilized the two verified datasets: the {Aggregated Benchmark} and the  {Reentrancy Scenarios Dataset (RSD)}.
For tool analysis, we adopted tools from \Cref{tab:tools}, and used SmartBugs 2.0 \cite{diAngeloEtAl2023ASE} to collect the results needed to assess reentrancy detection performance. 

All non-LLM models were evaluated through 3-fold cross-validation. Traditional ML models used \texttt{scikit-learn} v1.6.0 with TF-IDF features (max 2048) derived from Solidity code, optimizing hyperparameters via a grid search. DL models were implemented with \texttt{PyTorch} v2.4.1 and \texttt{transformers} v4.47.0, including a BiLSTM (hidden size 128) model using CodeBERT embeddings and a fine-tuned version of CodeBERT (\texttt{microsoft/codebert-base}), trained for up to 20 epochs with a learning rate of \(1 \times 10^{-3}\) and a batch size of 32, guided by a fixed validation set for early stopping. For the RSD, each training set fold was augmented with the entire Aggregated Benchmark to improve learning.

LLMs were evaluated zero-shot with a temperature of 0 
\chadded{to ensure deterministic outputs and reproducibility across runs, following empirical evaluations of LLM-based program analysis \cite{ouyang2025empirical}.}
\chadded{Beyond specifying the output format, the prompt represents a less formal version of the labeling procedure described in \Cref{sec:procedure}, defining external calls, state modifications, and reentrancy guards, and guiding the model to recognize complex scenarios such as cross-function and cross-contract reentrancy.}
Full prompt structures are available in our code repository\footnote{\chadded{Relative path from repository root: \url{src/llms/prompts.py}}}. 

Commercial LLMs were queried via vendor APIs from April 1 to May 20, 2025. Model snapshots used include \texttt{gpt-4o-2024-08-06}, \texttt{gpt-4.1-2025-04-14}, \texttt{o3-mini-2025-01-31}, \texttt{o4-mini-2025-04-16}; \texttt{gemini-2.0-flash-001}, and \texttt{gemini-2.5-flash-preview-05-20}. Explanations from \texttt{o3-mini} for 146 contracts in the Aggregated Benchmark were rated by three of the authors in a within-subject, randomized design, on a 5-point Likert scale. This evaluation was complemented by \texttt{o4-mini}'s analysis of all \texttt{o3-mini} explanations. Alongside commercial LLMs, we experimented with open-source models queried through their latest snapshots (December 2025) pulled from the Ollama repositories. A preliminary analysis of model language generation performance by size showed that small models (fewer than 10B parameters) were inadequate for the task at hand. Namely, we experimented with Gemma3:4b and Llama3.1:8b: they both struggled to generate structured output in a parsable JSON format, and even when they succeeded, their classification accuracy was underwhelming (with a strong bias towards classifying all contracts as safe). Thus, we shifted our focus towards the largest models that could fit consumer hardware (all local experiments were conducted on an NVIDIA GTX 5080 GPU). We selected three state-of-the-art mid-size LLMs for which we report accuracy performance: gpt-oss:20b, qwen3:14b, phi4:14b.

\subsection{Evaluation metrics}

The results derived from our two datasets (Table~\ref{tab:accuracy_combined}) highlight distinct capabilities and have important implications for reentrancy detection. For each row, in addition to Accuracy and F1 score, we report Precision and Recall, as these represent two critical aspects of our analysis.
Given a detector's predictions, let $\mathrm{TP}$, $\mathrm{FP}$, $\mathrm{TN}$, and $\mathrm{FN}$ denote true positives, false positives, true negatives, and false negatives, respectively. We compute:

{
\small
\begin{equation}
\mathrm{Accuracy} = \frac{\mathrm{TP} + \mathrm{TN}}{\mathrm{TP} + \mathrm{TN} + \mathrm{FP} + \mathrm{FN}}
\end{equation}

\begin{equation}
\mathrm{Precision} = \frac{\mathrm{TP}}{\mathrm{TP} + \mathrm{FP}}
\end{equation}

\begin{equation}
\mathrm{Recall} = \frac{\mathrm{TP}}{\mathrm{TP} + \mathrm{FN}}
\end{equation}

\begin{equation}
\mathrm{F1_{score}} = \frac{2 \cdot \mathrm{Precision} \cdot \mathrm{Recall}}{\mathrm{Precision} + \mathrm{Recall}}
\end{equation}
}


Low precision indicates a high number of false positives, a common criticism of static and formal-methods-based tools. Low recall, on the other hand, corresponds to a high number of false negatives—something a vulnerability detector should aim to minimize.
As the last metric, we report the error rate. For each tool and each contract, the detection output is considered an error/fail if the tool exited with an error code and did not provide any vulnerability in the list of possible vulnerabilities. While we initially discarded all outputs with exit codes different from 0, we observed that tool performance decreased drastically, so we applied this mitigation.

\section{Results}
\label{sec:results}

We present the results of our experiments on both the Aggregated Benchmark and the RSD in \Cref{tab:accuracy_combined}, divided into horizontal groups.
The first group includes detection tools and analyzers based on formal methods; the second group presents machine learning techniques and vanilla networks; the third group gathers commercial LLMs (called by API) as well as models installed locally (separated by a thin horizontal line). For each group and for each column, we underline the best result for the group, and mark the best result overall (of the whole column) in bold. For machine learning techniques, we also report the standard deviation, while the error is missing from both the machine learning and LLMs categories because it does not apply (the models always provide an answer).

\begin{table*}[t]
    \centering
    \caption{Model Performance Metrics on two datasets: \textit{Aggregated Benchmark} and \textit{RSD}.}
    \label{tab:accuracy_combined}
    \renewcommand{\arraystretch}{1.1} 

    \resizebox{0.95\textwidth}{!}{
    \begin{tabular}{|c|l | ccccr | ccccr |}
    \hline 
    & & 
\multicolumn{5}{c|}{{\normalsize\textbf{\textit{\rule{0pt}{1.2em}Aggregated Benchmark}}}} 
& \multicolumn{5}{c|}{{\normalsize\textbf{\textit{\rule{0pt}{1.2em}Reentrancy Scenarios Dataset (RSD)}}}} 
\\[6pt]

        & \textbf{Tool/Model} & \textbf{Accuracy} & \textbf{Precision} & \textbf{Recall} & \textbf{F1 Score} &  \textbf{Error (\%)}& \textbf{Accuracy} & \textbf{Precision} & \textbf{Recall} & \textbf{F1 Score} & \textbf{Error (\%)} \\
        \hline
        \multirow{10}{*}{\rotatebox{90}{\textbf{Analyzers}}} &
         Aderyn & - & - & - & - & 100.00 & 0.66 & 0.62 & 0.79 & 0.70 & \underline{0.00}  \\
        ~ & CCC & 0.89 & 0.91 & 0.66 & 0.76 & 25.46 & 0.58 & \underline{0.74} & 0.24 & 0.36 & \underline{0.00}  \\ 
        ~ & Confuzzius & 0.88 & 0.95 & 0.62 & 0.75 & 15.97 & 0.58 & 0.57 & 0.61 & 0.59 & 12.59  \\ 
        ~ & Conkas & 0.84 & 0.67 & 0.78 & 0.72 & 18.18 & - & - & - & - & 100.00  \\
        ~ & Mythril & 0.86 & 0.84 & 0.48 & 0.61 & 7.22 & 0.67 & 0.67 & 0.66 & 0.67 & \underline{0.00}  \\
        ~ & Oyente+ & 0.91 & 0.92 & 0.75 & 0.83 & \underline{0.00} & 0.60 & 0.72 & 0.32 & 0.45 & \underline{0.00}  \\ 
        ~ & Sailfish & 0.83 & 0.93 & 0.42 & 0.58 & \underline{0.00} & - & - & - & - & 100.00  \\ 
        ~ & Securify & 0.88 & \underline{\textbf{1.00}} & 0.48 & 0.65 & 25.54 & - & - & - & - & 100.00  \\ 
        ~ & Sfuzz & 0.76 & 0.71 & 0.24 & 0.36 & 45.14 & - & - & - & - & 100.00  \\ 
        ~ & Slither & \underline{0.95} & 0.88 & \underline{0.95} & \underline{0.92} & \underline{0.00} & \underline{0.74} & 0.71 & \underline{0.82} & \underline{0.76} & \underline{0.00}  \\ 
        ~ & Solhint & 0.81 & 0.90 & 0.36 & 0.51 & 0.69 & 0.55 & 0.59 & 0.28 & 0.38 & \underline{0.00}  \\ 
        & Vandal & 0.74 & 0.66 & 0.52 & 0.58 & 68.72 & 0.52 & 0.57 & 0.25 & 0.34 & 78.24 \\
        \hline
        \hline
        \multirow{10}{*}{\rotatebox{90}{\textbf{Machine Learning}}} &
        gradient\_boosting & $0.90 \pm 0.04$ & $0.87 \pm 0.04$ &  $0.76 \pm 0.11$ & $0.81 \pm 0.05$          &- & $0.62 \pm 0.02$ & $0.63 \pm 0.03$ & $0.62 \pm 0.02$ & $0.61 \pm 0.02$ & -\\
        & gnb & $0.82 \pm 0.05$ & $0.70 \pm 0.06$ & $0.65 \pm 0.12$ & $0.67 \pm 0.07$                         &- & $0.58 \pm 0.03$ & $0.71 \pm 0.05$ & $0.58 \pm 0.03$ & $0.51 \pm 0.07$ & -\\
        &  knn & $0.88 \pm 0.01$ & $0.85 \pm 0.13$ & $0.74 \pm 0.13$ & $0.78 \pm 0.01$                         &- & $0.62 \pm 0.01$ & $0.62 \pm 0.01$ & $0.62 \pm 0.01$ & $0.62 \pm 0.01$ & -\\
        &  logistic\_regression & $0.80 \pm 0.06$ & $\underline{0.93} \pm 0.12$ & $0.33 \pm 0.12$ & $0.47 \pm 0.12$        &- & $0.49 \pm 0.01$ & $0.33 \pm 0.12$ & $0.49 \pm 0.01$ & $0.35 \pm 0.03$ & -\\
        &  random\_forest & $0.90 \pm 0.04$ & $0.90 \pm 0.07$ & $0.73 \pm 0.13$ & $0.80 \pm 0.07$              &- & $0.63 \pm 0.03$ & $0.64 \pm 0.03$ & $0.63 \pm 0.03$ & $0.62 \pm 0.04$ & -\\
        & svm & $0.84 \pm 0.04$ & $\underline{0.93} \pm 0.12$ & $0.51 \pm 0.16$ & $0.64 \pm 0.09$                         &- & $0.50 \pm 0.00$ & $0.33 \pm 0.01$ & $0.50 \pm 0.00$ & $0.34 \pm 0.02$ & -\\
        & xgboost & $0.91 \pm 0.02$ & $0.88 \pm 0.08$ & $0.79 \pm 0.10$ & $0.83 \pm 0.02$                     &- & $\underline{0.72} \pm 0.03$ & $\underline{0.72} \pm 0.03$ & $\underline{0.72} \pm 0.03$ & $\underline{0.72} \pm 0.03$ & -\\
        & codebert & $0.90 \pm 0.13$ & $0.82 \pm 0.24$ & $\textbf{\underline{0.96}} \pm 0.03$ & $0.87 \pm 0.14$               &- & $0.65 \pm 0.06$ & $0.67 \pm 0.07$ & $0.65 \pm 0.06$ & $0.64 \pm 0.06$ & -\\
        & lstm & $0.86 \pm 0.12$ & $0.76 \pm 0.17$ & $\textbf{\underline{0.96}} \pm 0.03$ & $0.83 \pm 0.11$               &- & $0.60 \pm 0.05$ & $0.62 \pm 0.06$ & $0.60 \pm 0.05$ & $0.59 \pm 0.05$ & -\\
        & ffnn & $\underline{\textbf{0.96}} \pm 0.01$ & $0.85 \pm 0.03$ & $0.89 \pm 0.03$ & $\underline{0.86} \pm 0.02$               &- & $0.61 \pm 0.03$ & $0.62 \pm 0.03$ & $0.61 \pm 0.03$ & $0.61 \pm 0.03$ & -\\
        \hline
        \hline
        
        \multirow{10}{*}{\rotatebox{90}{\textbf{LLMs}}} & gemini-2.0-flash & $0.87$ & $0.87$ & $0.84$ & $0.85$ &- & $0.51$ & $0.52$ & $0.52$ & $0.49$ & -\\
        & gemini-2.5-flash & $0.83$ & $0.85$ & $0.82$ & $0.83$ &- & $0.58$ & $0.66$ & $0.58$ & $0.53$ & -\\
        & gpt-4o & $0.85$ & $0.89$ & $0.85$ & $0.86$ &- & $0.53$ & $0.53$ & $0.53$ & $0.51$ & -\\
        & gpt-4.1 & $0.87$ & $0.90$ & $0.87$ & $0.88$ &- & $0.63$ & $0.63$ & $0.63$ & $0.63$ & -\\
        & o3-mini & $\underline{\textbf{0.96}}$ & $\underline{0.96}$ & $\underline{\textbf{0.96}}$ & $\underline{\textbf{0.96}}$ &- & $0.79$ & $0.80$ & $0.79$ & $0.79$ & -\\
        & o4-mini & $0.94$ & $0.94$ & $0.94$ & $0.93$ & - & ${0.82}$ & ${0.83}$ & $\underline{\textbf{0.82}}$ & $\underline{\textbf{0.82}}$ & -\\

        &         gpt-oss & 0.94 & \underline{0.96} & 0.82 & 0.88 & - & \underline{\textbf{0.87}} & \underline{\textbf{0.98}} & 0.75 & 0.85 & -  \\ 
        &qwen3 & 0.91 & 0.78 & 0.93 & 0.85 & - & 0.68 & 0.66 & 0.73 & 0.69 & -  \\ 
        &phi4 & 0.80 & 0.58 & \underline{\textbf{0.96}} & 0.72 & - & 0.56 & 0.55 & 0.65 & 0.60 & - \\ 
        \hline
    \end{tabular}}

\end{table*}

\subsection{\textbf{RQ1:} How reliable and robust are existing reentrancy detection tools when applied to modern Solidity codebases and evaluated under stress-testing scenarios, given their lack of maintenance and outdated compiler support?}
To answer this question, we analyze the error rates reported in \Cref{tab:accuracy_combined}. Among the 11 tools running on the Aggregated Benchmark, only 7 can process contracts in the RSD dataset. Of these, just 5 complete the analysis without failures. A notable exception is Aderyn, a recently developed tool that focuses exclusively on newer Solidity versions. CCC and Oyente+ also show broader version support. Across the RSD, tools tend to either fully support or entirely fail on contracts (with Confuzzius and Vandal being the only partial cases).

Older tools that remain actively maintained — such as Mythril, Slither, and Solhint — still encounter a substantial number of errors even in the Aggregated Benchmark. This can be attributed to the wide range of compiler versions in the dataset and the numerous syntax and semantic changes introduced within Solidity 0.4.x and 0.5.x. Moreover, most analyzers operate at the bytecode (EVM opcode) level, where historical differences in EVM revisions further complicate compatibility.
High error rates also influence the interpretation of performance metrics. For instance, Securify reports 100\% precision in the RSD dataset primarily because many contracts it fails to analyze are labeled as reentrant, inflating the metric. Overall, tool performance degrades noticeably when moving from the Aggregated Benchmark to the RSD, even when considering only analyzers that do not report errors.

\subsection{\textbf{RQ2:} How consistent are the warnings and vulnerability reports produced by different tools when applied to the same contracts?} 
\begin{figure*}[!ht]
    \centering
    
    \begin{subfigure}{\linewidth}
        \centering
        
        \label{fig:upsetplot:agg}
        \includegraphics[width=\linewidth]{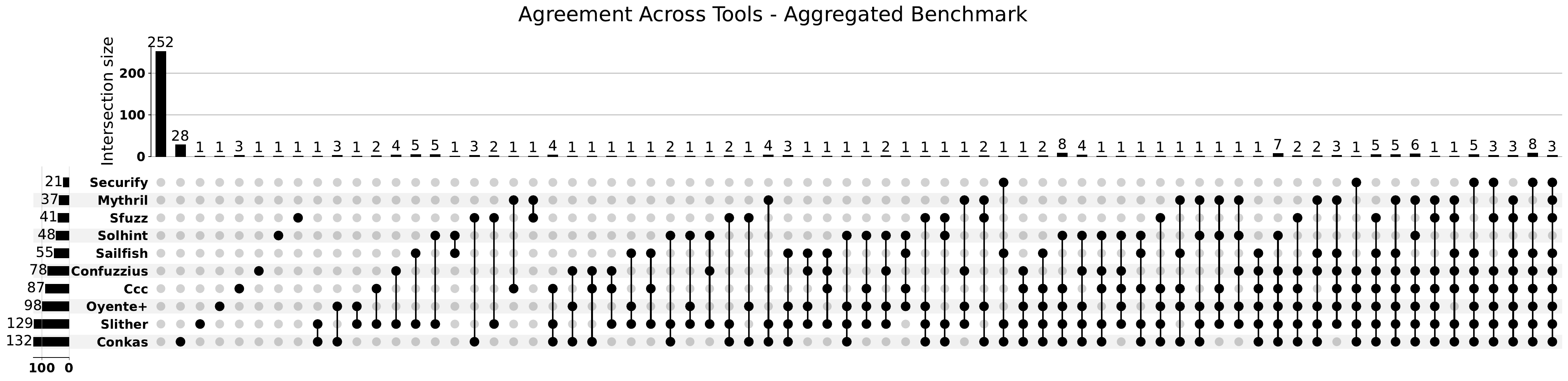}
    \end{subfigure}

    \vspace{3pt}

    \begin{subfigure}{0.55\linewidth}
        \centering

        \label{fig:upsetplot:rsd}
                \includegraphics[width=\linewidth]{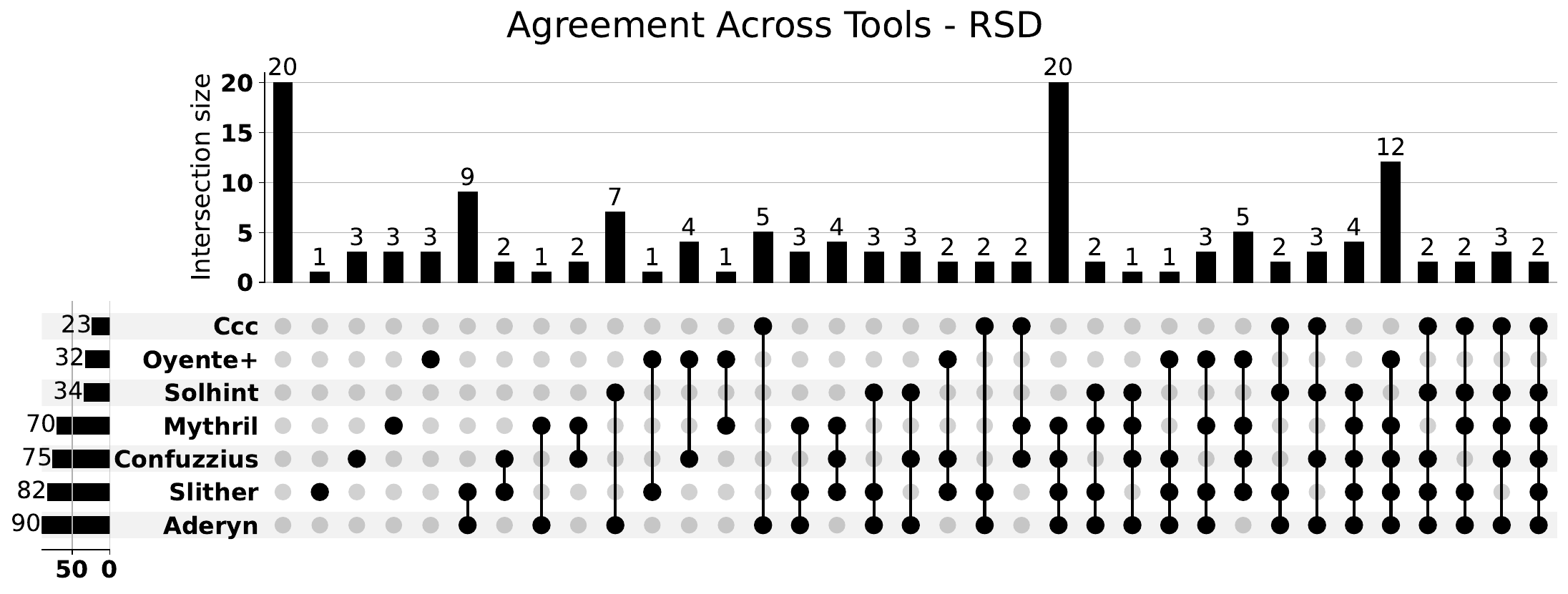}
    \end{subfigure}

    \caption{Comparison between upset plots for Aggregated Benchmark and RSD datasets. 
    \chadded{This type of plot serves as an alternative to Venn diagrams for visualizing intersections and overlaps among more than three sets. The histograms indicate the number of samples per row and column, while the dots show which tools flagged the contracts within each set.}}
    \label{fig:upsetplot}
\end{figure*}

\chadded{Figure}~\ref{fig:upsetplot} shows two upset plots that illustrate how much different tools agree when reporting reentrancy vulnerabilities. Each plot can be read as follows: the horizontal bars on the left indicate how many contracts each tool flagged, while the filled dots below the plot identify which tools are included in each intersection. A vertical bar is then drawn above every connected group of dots, quantifying how many contracts were jointly marked as reentrant by that specific combination of tools. Thus, the dots define the tool combination and the height of the bar shows the size of the agreement.

The plots reveal very limited agreement both in real-world scenarios (Aggregated Benchmark) and in MWEs (RSD).
For the top plot, all tools agree on the lack of reentrancy over 252 contracts, regardless of the ground truth. On the opposite side, though, if we consider the agreement among at least 6 tools, we count only 78 contracts out of 122 labeled as reentrant. 
For the RSD the agreement is a bit stronger, probably due to the reduced number of tools supporting Solidity v0.8 and the cardinality of the dataset.
In particular, there is a similarity in the detection of Mythril, Confuzzius, Slither, and Aderyn (20 contracts).

These results suggest that automated tools tend to converge only when reentrancy cases are simple and well-defined, while more realistic implementations still lead to substantial disagreement across analyzers.
For completeness, we report which categories of reentrant contracts in the RSD are most frequently misclassified by the four top-performing detectors:

\vspace{1ex}
\noindent\resizebox{\columnwidth}{!}{
    \rowcolors{1}{white}{lightgray!30}
\centering
    \begin{tabular}{l|p{4.5cm}|p{4.8cm}}
    \hline
        ~ & \textbf{Code} & \textbf{Variants}  \\ \hline
        \textbf{Aderyn} & 00, 03, 04, 12, 14 & Fold, Inline, Underflow, Mod  \\ 
        \textbf{Mythril} & 00, 01, 04, 09, 12, 13, 14, 15 & Error, Mod, Fold, Staticcall  \\ 
        \textbf{Oyente+} & 00, 01, 02, 03, 04, 07, 08, 09, 13, 14, 15 & Inline, Fold, Underflow, Unchecked, Emit, Mod, Staticcall  \\ 
        \textbf{Slither} & 00, 09, 14, 15 & Inline, Mod, Staticcall \\ \hline
    \end{tabular}
}
\vspace{1ex}

\chadded{Oyente+ misses several cases due to unsupported post-0.8 features, while Mythril and Slither overlook some baseline scenarios (code 00, see \Cref{tab:scenarios} and \Cref{sec:rsd}) depending on how CEI-breaking is interpreted (e.g., whether \texttt{error} and \texttt{emit} count as effects). Static calls are often deemed harmless, whereas our procedure labels them as exploitable via read-only reentrancy \cite{sachinoglou2023readonly}. Aderyn, operating on source code, misses subtler patterns such as function folding and nested modifiers; cross-function cases with mutex-style guards (code 04) are also challenging. Delegatecalls (code 14) are not always conservatively treated as malicious. Overall, discrepancies persist due to divergent reentrancy interpretations and mismatches between published descriptions and actual implementations.}

\subsection{\textbf{RQ3:} Do general-purpose Large Language Models (LLMs) demonstrate higher \chadded{dependability} and lower maintenance cost compared to static analysis and ML-based detectors?}
As shown in \Cref{tab:accuracy_combined}, LLMs achieve the best overall performance across both datasets. In the Aggregated Benchmark, \texttt{o3-mini} nearly reaches perfect accuracy, surpassing both the strongest static analyzer (Slither) and the top ML model (FFNN). In the RSD dataset, \texttt{gpt-oss} achieves the highest scores among all evaluated approaches, confirming that high accuracy is attainable even with fully local models.

LLMs also provide stronger robustness: unlike static analyzers, they do not fail due to unsupported compiler versions or outdated vulnerability rules, resulting in zero reported errors in both datasets. This reliability stems from their ability to reason directly over source semantics rather than relying on hand-crafted bytecode patterns.

\chadded{Lastly, maintenance costs are negligible: improvements in foundation models generally improve detection performance without requiring rule updates or model retraining, although newer models may not outperform earlier ones in all cases.}
Notably, the commercial LLMs evaluated in this study (GPT-4.1, GPT-4o, o4-mini) have already been superseded by the more capable GPT-5.1 family. Our results therefore represent a conservative lower bound on current LLM performance, further reinforcing the widening robustness gap relative to static analyzers.


\chadded{Overall, LLMs introduce a tradeoff between determinism and robustness: traditional analyzers provide reproducible rule-based reasoning but require continuous maintenance, whereas LLMs sacrifice strict determinism while remaining robust to language evolution and unseen attack patterns. As foundation models improve without manual updates, this tradeoff is likely to increasingly favor LLM-based approaches.}



\subsection{\textbf{RQ4:} Can LLM-generated explanations assist human experts in identifying mislabeled or ambiguous cases, thus improving dataset quality?}

LLMs played a central role in validating the annotations of our two datasets. Traditional analysis tools often produce logs that are difficult to interpret, while LLMs return concise and readable explanations that identify the vulnerable function and the relevant code location. This clear communication is particularly important in security analysis, where human oversight is essential for trust. During our verification process, reviewing the explanations not only confirmed detection outcomes but also revealed multiple annotation issues.

In particular, LLM-generated explanations allowed us to identify \chadded{9 contracts in the Aggregated dataset} that were initially mislabeled due to subtle or ambiguous reentrancy behavior. 
By comparison, Slither detected only 3 mislabeled cases in the RSD dataset, and understanding its reasoning required significantly more time than analyzing the explanations provided by \texttt{o3-mini}. This demonstrates the practical value of LLMs in iterative dataset improvement and supports their use as complementary tools for security auditing.

After dataset construction, the three authors evaluated the goodness of the explanation for contracts misclassified by \texttt{o3-mini}.
\chadded{Given the complexity and the large number of explanations to evaluate (more than \num{560000}, considering all models and all contracts), we used an LLM-as-a-judge approach for the rest of the responses.}

\vspace{1ex}
\noindent\resizebox{\columnwidth}{!}{
    {
    \centering
    \rowcolors{1}{white}{lightgray!30}
    \begin{tabular}{l|c|c}
    \hline
    \textbf{Metric} &
    \textbf{\begin{tabular}{c}Human Evaluation \\(Misclassified)\end{tabular}} &
    \textbf{\begin{tabular}{c}LLM Evaluation\\(Full Set)\end{tabular}} \\
    \hline
Correctness     & $4.38 \,\pm\, 0.94$ & $4.46 \,\pm\, 0.91$ \\
Informativeness & $4.14 \,\pm\, 0.70$ & $3.32 \,\pm\, 0.81$ \\
Pertinence      & $4.37 \,\pm\, 0.87$ & $4.81 \,\pm\, 0.57$ \\

    \hline
    \end{tabular}
    }
}

\medskip\noindent
Human reviewers scored explanations based on the correctness of the explanation, if it was informative enough to understand why a contract was reentrant or not, and if the answer was pertinent to reentrancy, showing generally positive results. The automatic evaluation, conducted over the entire dataset, confirmed similar correctness and even higher pertinence, though informativeness was scored lower, possibly due to more demanding expectations regarding depth or novelty. Qualitative feedback indicates that the most effective explanations pointed to specific functions or critical state changes within the code, while a minority of explanations lacked useful detail.

Overall, these observations show that LLMs offer more helpful and actionable insights than current static analyzers and can support human experts in verifying challenging examples. Their involvement leads to improved dataset reliability and contributes to more effective benchmarking in smart contract security.

\section{Threats to Validity}

\textbf{Internal validity.}
Manual relabeling may still contain residual errors, despite three-reviewer consensus and LLM-assisted cross-checks. However, our results show a \emph{systematic lack of agreement} among tools themselves (Table~\ref{tab:accuracy_combined}, Figure~\ref{fig:upsetplot}), including clear-cut cases. These divergences persist even if annotation noise is assumed, indicating that detector inconsistencies stem primarily from tool heuristics rather than ground truth quality.
Execution outcomes also depend on the SmartBugs framework 2.0.15 used to run most tools, and on legacy compiler support when analyzing modern Solidity code. We mitigated instability by executing tools in a controlled environment and repeating failing runs, but partial outputs may still influence individual results. 

\chadded{Additionally, using LLMs to assist humans in the manual labeling process may introduce subtle confirmation or anchoring biases, potentially influencing how borderline or ambiguous cases are interpreted. If the labeling process is partially shaped by model-generated explanations or suggestions, this could inadvertently align the ground truth with patterns more easily recognized by similar models. Such an effect may partly explain why \texttt{o3-mini} achieves particularly high performance in our evaluation.}

\textbf{External validity.}
The Aggregated Benchmark includes real-world contracts but is not statistically representative of the entire Ethereum ecosystem, while the RSD intentionally focuses on semantic stress tests rather than prevalence. Thus, our findings should not be interpreted as estimates of ecosystem-wide vulnerability rates; rather, they provide evidence of reliability and robustness gaps in current detection tools. Broader studies will be needed to assess how these conclusions extend across more diverse contract populations and execution environments.

\section{Related Work}

\chadded{Empirical evaluations of smart contract analyzers commonly rely on manually labeled datasets spanning multiple vulnerability classes.
Frameworks such as SmartBugs~\cite{ferreira2020smartbugs} and subsequent curated benchmarks evaluate static and dynamic analyzers using heterogeneous vulnerability corpora, while recent studies provide large manually annotated datasets with line-level labels for several vulnerability types~\cite{salzano2025empirical}. 
Although these efforts improve benchmarking reproducibility, reentrancy represents only one category among many, resulting in limited semantic coverage of this specific vulnerability, especially when vulnerabilities are injected \cite{ghaleb2020effective,huanggai}.}

\chadded{A second line of work focuses explicitly on reentrancy detection.
Prior studies and tools analyze reentrancy patterns or propose specialized detection techniques~\cite{reentrancy_study,autoar2025,wu2024advscanner}. 
However, these works typically consider a restricted set of analyzers and do not explicitly characterize \emph{dependability under language evolution}, \ie whether detectors remain functional and reliable on modern Solidity (0.8+), where syntax and semantics have substantially changed.
Moreover, existing reentrancy-focused benchmarks do not combine \emph{minimal working examples} designed as controlled stress tests that isolate specific semantic corner cases (e.g., \texttt{delegatecall}, \texttt{staticcall}, proxies, folded calls) with a fine-grained taxonomy that goes beyond the traditional macroscopic categories (single-function, cross-function, cross-contract etc.), making it difficult to attribute failures to precise reentrancy mechanisms rather than to incidental contract complexity.}

\chadded{More recently, several works have explored the use of large language models for smart contract vulnerability detection~\cite{boi2024smart,chen2025chatgpt,wei2025advanced,sun2024gptscan}. 
These studies demonstrate promising capabilities of LLMs but evaluate multiple vulnerability classes simultaneously, without conducting a dedicated analysis of reentrancy semantics or robustness across evolving language features.
}

\chadded{In contrast, the present work concentrates exclusively on reentrancy and provides a unified dependability evaluation across static analyzers, ML-based approaches, and modern LLMs, grounded in two manually verified datasets designed to capture both real-world contracts and semantically challenging reentrancy scenarios.}

\section{Conclusion}
\label{sec:conclusions}

This work revisits the state of reentrancy detection from a dependability perspective. We introduce two manually verified datasets: the \textit{Aggregated Benchmark}, built from three prior human-labeled corpora, and the \textit{Reentrancy Scenarios Dataset (RSD)}, a stress-testing suite covering modern Solidity features and semantic corner cases. Using these datasets, we evaluate 12 traditional tools, 10 ML/DL models, and 9 LLMs.

Our findings expose a fractured and unreliable ecosystem.
Despite strong F1 scores for detectors such as Oyente+ (83\%) and Slither (92\%), the remaining tools perform extremely poorly, often failing with errors or producing no meaningful output. 
More concerning is the lack of consistency: on the Aggregated Benchmark, only 38 out of 122 reentrant contracts exhibit agreement among at least 6 tools, meaning that even a minimal majority consensus is reached for barely one third of the cases.
On the RSD, many tools fail outright due to outdated compiler support or simplistic syntactic heuristics. 
In contrast, multiple LLMs achieve above 85\% accuracy in zero-shot settings, outperforming all conventional detectors while requiring no maintenance. Their explanations also helped identify mislabeled cases, directly improving dataset reliability.

These results indicate a fundamental shift: as Solidity evolves, static tools are no longer dependable ground truth for security pipelines, while LLMs already provide more robust and semantically grounded analysis. Still, several challenges remain. Our datasets, though carefully curated, cover a limited fraction of deployed contract diversity. Scaling human-in-the-loop validation of LLM explanations and investigating more advanced prompting or fine-tuning strategies represent promising directions. Moreover, hybrid approaches that combine the formal guarantees of static analysis with the adaptability of LLMs may offer the best path toward dependable vulnerability detection in rapidly evolving smart contract ecosystems.

\section*{Data Availability Statement}
The data analyzed and discussed in this paper are openly available in a repository together with the supplementary material.
The link to the repository is provided in anonymized form for the review process: \url{https://anonymous.4open.science/r/reentrancy-detection-benchmarks-A1D7}.



\bibliographystyle{plain}
\bibliography{bibliography}

@inproceedings{luu2016oyente,
	title        = {Making smart contracts smarter},
	author       = {Luu, Loi and Chu, Duc-Hiep and Olickel, Hrishi and Saxena, Prateek and Hobor, Aquinas},
	year         = 2016,
	month        = 10,
	journal      = {Proceedings of the 2022 ACM SIGSAC Conference on Computer and Communications Security},
	booktitle    = {Proceedings of the ACM SIGSAC Conference on Computer and Communications Security (CCS)},
	publisher    = {ACM},
	pages        = {254--269},
	url          = {https://doi.org/10.1145/2976749.2978309},
	editor       = {Weippl, Edgar R. and Katzenbeisser, Stefan and Kruegel, Christopher and Myers, Andrew C. and Halevi, Shai}
}

@inproceedings{autoar2025,
	title        = {Silence False Alarms: Identifying Anti-Reentrancy Patterns on {Ethereum} to Refine Smart Contract Reentrancy Detection},
	author       = {Gao, Jun and Fu, Yu and Chen, Zhi and Ye, Defang and Qian, Peng and Sun, Kun},
	year         = 2025,
	booktitle    = {Proceedings of the Network and Distributed System Security Symposium (NDSS)}
}

@inproceedings{diAngeloEtAl2023ASE,
	title        = {{SmartBugs} 2.0: An Execution Framework for Weakness Detection in Ethereum Smart Contracts},
	author       = {di Angelo, Monika and Durieux, Thomas and Ferreira, Jo{\~a}o F. and Salzer, Gernot},
	year         = 2023,
	month        = 9,
	journal      = {2021 36th IEEE/ACM International Conference on Automated Software Engineering (ASE)},
	booktitle    = {38th IEEE/ACM International Conference on Automated Software Engineering (ASE)},
	publisher    = {IEEE Computer Society},
	pages        = {2102--2105},
	doi          = {10.1109/ASE56229.2023.00060},
	url          = {https://repository.tudelft.nl/file/File_7ea75f0a-3db5-47a2-ad85-118be1e89eb9}
}

@inproceedings{luu2016making,
	title        = {Making smart contracts smarter},
	author       = {Luu, Loi and Chu, Duc-Hiep and Olickel, Hrishi and Saxena, Prateek and Hobor, Aquinas},
	year         = 2016,
	month        = 10,
	journal      = {Proceedings of the 2022 ACM SIGSAC Conference on Computer and Communications Security},
	booktitle    = {Proceedings of the 2016 ACM SIGSAC conference on computer and communications security},
	publisher    = {ACM},
	pages        = {254--269},
	editor       = {Weippl, Edgar R. and Katzenbeisser, Stefan and Kruegel, Christopher and Myers, Andrew C. and Halevi, Shai}
}

@inproceedings{feist2019slither,
	title        = {Slither: a static analysis framework for smart contracts},
	author       = {Feist, Josselin and Grieco, Gustavo and Groce, Alex},
	year         = 2019,
	month        = 5,
	journal      = {International Workshop on Emerging Trends in Software Engineering for Blockchain},
	booktitle    = {2019 IEEE/ACM 2nd International Workshop on Emerging Trends in Software Engineering for Blockchain (WETSEB)},
	publisher    = {IEEE / ACM},
	pages        = {8--15},
	doi          = {10.1109/wetseb.2019.00008},
	url          = {https://arxiv.org/pdf/1908.09878},
	organization = {IEEE}
}

@inproceedings{torres2021confuzzius,
	title        = {Confuzzius: A data dependency-aware hybrid fuzzer for smart contracts},
	author       = {Torres, Christof Ferreira and Iannillo, Antonio Ken and Gervais, Arthur and State, Radu},
	year         = 2021,
	month        = 9,
	booktitle    = {2021 IEEE European Symposium on Security and Privacy (EuroS\&P)},
	publisher    = {IEEE},
	pages        = {103--119},
	doi          = {10.1109/eurosp51992.2021.00018},
	url          = {https://discovery.ucl.ac.uk/id/eprint/10182330/1/2005.12156.pdf},
	organization = {IEEE}
}

@inproceedings{yang2024uncover,
	title        = {Uncover the premeditated attacks: Detecting exploitable reentrancy vulnerabilities by identifying attacker contracts},
	author       = {Yang, Shuo and Chen, Jiachi and Huang, Mingyuan and Zheng, Zibin and Huang, Yuan},
	year         = 2024,
	month        = 4,
	journal      = {International Conference on Software Engineering},
	booktitle    = {Proceedings of the IEEE/ACM 46th International Conference on Software Engineering},
	publisher    = {ACM},
	pages        = {1--12},
	url          = {https://arxiv.org/pdf/2403.19112}
}

@inproceedings{tsankov2018securify,
	title        = {Securify: Practical security analysis of smart contracts},
	author       = {Tsankov, Petar and Dan, Andrei and Drachsler-Cohen, Dana and Gervais, Arthur and Buenzli, Florian and Vechev, Martin},
	year         = 2018,
	month        = 6,
	journal      = {arXiv (Cornell University)},
	booktitle    = {Proceedings of the 2018 ACM SIGSAC conference on computer and communications security},
	publisher    = {Cornell University},
	volume       = {abs/1806.01143},
	pages        = {67--82},
	doi          = {10.48550/arxiv.1806.01143},
	url          = {https://arxiv.org/abs/1806.01143},
	editor       = {Lie, David and Mannan, Mohammad and Backes, Michael and Wang, Xiaofeng}
}

@inbook{cgt,
	title        = {Consolidation of Ground Truth Sets for Weakness Detection in Smart Contracts},
	author       = {Di Angelo, Monika and Salzer, Gernot},
	year         = 2023,
	month        = 12,
	booktitle    = {Financial Cryptography and Data Security. FC 2023 International Workshops - Voting, CoDecFin, DeFi, WTSC, Bol, Brač, Croatia, May 5, 2023, Revised Selected Papers},
	publisher    = {Springer Science+Business Media},
	volume       = 13953,
	pages        = {439--455},
	doi          = {10.1007/978-3-031-48806-1_28},
	isbn         = {978-3-031-48805-4},
	issn         = {0302-9743},
	url          = {https://doi.org/10.1007/978-3-031-48806-1_28},
	editor       = {Essex, Aleksander and Matsuo, Shin'ichiro and Kulyk, Oksana and Gudgeon, Lewis and Klages-Mundt, Ariah and Perez, Daniel and Werner, Sam and Bracciali, Andrea and Goodell, Geoff}
}

@inproceedings{reentrancy_study,
	title        = {Turn the Rudder: A Beacon of Reentrancy Detection for Smart Contracts on Ethereum},
	author       = {Zheng, Zibin and Zhang, Neng and Su, Jianzhong and Zhong, Zhijie and Ye, Mingxi and Chen, Jiachi},
	year         = 2023,
	month        = 5,
	journal      = {International Conference on Software Engineering},
	booktitle    = {Proceedings of the 45th International Conference on Software Engineering},
	location     = {Melbourne, Victoria, Australia},
	publisher    = {IEEE Press},
	series       = {ICSE '23},
	pages        = {295–306},
	doi          = {10.1109/ICSE48619.2023.00036},
	isbn         = 9781665457019,
	url          = {https://doi.org/10.1109/ICSE48619.2023.00036},
	numpages     = 12
}

@article{turchin1986concept,
	title        = {The concept of a supercompiler},
	author       = {Turchin, Valentin F},
	year         = 1986,
	month        = 6,
	journal      = {ACM Transactions on Programming Languages and Systems (TOPLAS)},
	publisher    = {ACM New York, NY, USA},
	volume       = 8,
	number       = 3,
	pages        = {292--325},
	issn         = {0164-0925}
}

@inproceedings{nguyen2020sfuzz,
	title        = {sfuzz: An efficient adaptive fuzzer for solidity smart contracts},
	author       = {Nguyen, Tai D and Pham, Long H and Sun, Jun and Lin, Yun and Minh, Quang Tran},
	year         = 2020,
	month        = 4,
	journal      = {arXiv (Cornell University)},
	booktitle    = {Proceedings of the ACM/IEEE 42nd International Conference on Software Engineering},
	publisher    = {Cornell University},
	volume       = {abs/2004.08563},
	pages        = {778--788},
	doi          = {10.48550/arxiv.2004.08563},
	url          = {https://arxiv.org/abs/2004.08563},
	editor       = {Rothermel, Gregg and Bae, Doo-Hwan}
}

@article{ressi2024vulnerability,
	title        = {Vulnerability Detection in Ethereum Smart Contracts via Machine Learning: A Qualitative Analysis},
	author       = {Ressi, Dalila and Span{\`o}, Alvise and Benetollo, Lorenzo and Piazza, Carla and Bugliesi, Michele and Rossi, Sabina},
	year         = 2024,
	month        = 7,
	journal      = {arXiv preprint arXiv:2407.18639},
	publisher    = {Cornell University},
	volume       = {abs/2407.18639},
	doi          = {10.48550/arxiv.2407.18639},
	url          = {https://arxiv.org/pdf/2407.18639}
}

@inproceedings{DurieuxEtAl2020ICSE,
	title        = {Empirical Review of Automated Analysis Tools on 47,587 {Ethereum} Smart Contracts},
	author       = {Durieux, Thomas and Ferreira, Jo{\~a}o F. and Abreu, Rui and Cruz, Pedro},
	year         = 2020,
	month        = 6,
	booktitle    = {Proceedings of the ACM/IEEE 42nd International conference on software engineering},
	publisher    = {ACM},
	pages        = {530--541},
	url          = {https://arxiv.org/pdf/1910.10601},
	editor       = {Rothermel, Gregg and Bae, Doo-Hwan}
}

@article{wang2020contractward,
	title        = {Contractward: Automated vulnerability detection models for ethereum smart contracts},
	author       = {Wang, Wei and Song, Jingjing and Xu, Guangquan and Li, Yidong and Wang, Hao and Su, Chunhua},
	year         = 2020,
	month        = 1,
	journal      = {IEEE Transactions on Network Science and Engineering},
	publisher    = {IEEE},
	volume       = 8,
	number       = 2,
	pages        = {1133--1144},
	doi          = {10.1109/tnse.2020.2968505},
	issn         = {2327-4697},
	url          = {https://ntnuopen.ntnu.no/ntnu-xmlui/bitstream/11250/2639054/1/ContractWard.pdf}
}

@misc{smartbugsConkasGithub,
	title        = {{G}it{H}ub - smartbugs/conkas at \#4e0f256 --- github.com},
	author       = {},
	year         = {},
	note         = {Accessed November 2025},
	howpublished = {\url{https://github.com/smartbugs/conkas}}
}

@misc{mythrilGithub,
	title        = {{G}it{H}ub - {C}onsen{S}ys{D}iligence/mythril at --v0.24.8 - github.com},
	author       = {},
	year         = {},
	note         = {Accessed November 2025},
	howpublished = {\url{https://github.com/ConsenSysDiligence/mythril}}
}

@inproceedings{ferreira2020smartbugs,
	title        = {Smartbugs: A framework to analyze solidity smart contracts},
	author       = {Ferreira, Jo{\~a}o F and Cruz, Pedro and Durieux, Thomas and Abreu, Rui},
	year         = 2020,
	month        = 7,
	journal      = {arXiv (Cornell University)},
	booktitle    = {Proceedings of the 35th IEEE/ACM international conference on automated software engineering},
	publisher    = {Cornell University},
	pages        = {1349--1352},
	doi          = {10.48550/arxiv.2007.04771},
	url          = {https://arxiv.org/pdf/2007.04771}
}

@inproceedings{ghaleb2020effective,
	title        = {How effective are smart contract analysis tools? evaluating smart contract static analysis tools using bug injection},
	author       = {Ghaleb, Asem and Pattabiraman, Karthik},
	year         = 2020,
	month        = 7,
	booktitle    = {Proceedings of the 29th ACM SIGSOFT international symposium on software testing and analysis},
	publisher    = {ACM},
	pages        = {415--427},
	url          = {https://arxiv.org/pdf/2005.11613},
	editor       = {Khurshid, Sarfraz and Pasareanu, Corina S.}
}

@article{zhang2024smartreco,
	title        = {SmartReco: Detecting read-only reentrancy via fine-grained cross-DApp analysis},
	author       = {Zhang, Jingwen and Zheng, Zibin and Nan, Yuhong and Ye, Mingxi and Ning, Kaiwen and Zhang, Yu and Zhang, Weizhe},
	year         = 2024,
	month        = 9,
	journal      = {arXiv preprint arXiv:2409.18468},
	publisher    = {Cornell University},
	pages        = {2138--2150},
	doi          = {10.48550/arxiv.2409.18468},
	url          = {https://arxiv.org/pdf/2409.18468}
}

@article{burstall1977transformation,
	title        = {A Transformation System for Developing Recursive Programs},
	author       = {Burstall, R. M. and Darlington, John},
	year         = 1977,
	month        = 1,
	journal      = {Journal of the ACM},
	publisher    = {Association for Computing Machinery},
	volume       = 24,
	number       = 1,
	pages        = {44--67},
	doi          = {10.1145/321992.321996},
	issn         = {0004-5411}
}

@article{partsch1983program,
	title        = {Program transformation systems},
	author       = {Partsch, Helmuth and Steinbr{\"u}ggen, Ralf},
	year         = 1983,
	month        = 9,
	journal      = {ACM Computing Surveys (CSUR)},
	publisher    = {ACM New York, NY, USA},
	volume       = 15,
	number       = 3,
	pages        = {199--236},
	issn         = {0360-0300}
}

@inproceedings{perez2021smart,
	title        = {Smart contract vulnerabilities: Vulnerable does not imply exploited},
	author       = {Perez, Daniel and Livshits, Benjamin},
	year         = 2021,
	journal      = {USENIX Security Symposium},
	booktitle    = {30th USENIX Security Symposium (USENIX Security 21)},
	publisher    = {USENIX Association},
	pages        = {1325--1341},
	url          = {https://www.usenix.net/system/files/sec21summer_perez.pdf},
	editor       = {Bailey, Michael and Greenstadt, Rachel}
}

@inproceedings{schneidewind2020ethor,
	title        = {ethor: Practical and provably sound static analysis of ethereum smart contracts},
	author       = {Schneidewind, Clara and Grishchenko, Ilya and Scherer, Markus and Maffei, Matteo},
	year         = 2020,
	month        = 10,
	journal      = {Conference on Computer and Communications Security},
	booktitle    = {Proceedings of the 2020 ACM SIGSAC Conference on Computer and Communications Security},
	publisher    = {ACM},
	pages        = {621--640},
	url          = {https://arxiv.org/pdf/2005.06227},
	editor       = {Ligatti, Jay and Ou, Xinming and Katz, Jonathan and Vigna, Giovanni}
}

@inproceedings{rodler2019sereum,
	title        = {Sereum: Protecting Existing Smart Contracts Against Re-Entrancy Attacks},
	author       = {Michael Rodler and Wenting Li and Ghassan O. Karame and Lucas Davi},
	year         = 2019,
	booktitle    = {Proceedings of the Network and Distributed System Security Symposium (NDSS)},
	publisher    = {Internet Society},
	doi          = {10.14722/ndss.2019.23413},
	url          = {https://arxiv.org/abs/1812.05934},
	note         = {Runtime defense and taxonomy of advanced re-entrancy patterns; shows attacks beyond the DAO archetype, including those using method calls.}
}

@misc{sachinoglou2023readonly,
	title        = {Read-Only Reentrancy},
	author       = {Ioannis Sachinoglou},
	year         = 2023,
	note         = {“A novel attack class responsible for \$100M+ funds at risk”},
	howpublished = {ChainSecurity presentation/pdf}
}

@misc{aderynGithub,
	title        = {{G}it{H}ub - Cyfrin/aderyn at \#f4ce952
--- github.com},
	author       = {},
	year         = {},
	note         = {Accessed November 2025},
	howpublished = {\url{https://github.com/Cyfrin/aderyn}}
}

@inproceedings{bose2021sailfish,
	title        = {SAILFISH: Vetting Smart Contract State-Inconsistency Bugs in Seconds},
	author       = {Priyanka Bose and Dipanjan Das and Yanju Chen and Yu Feng and Christopher Kruegel and Giovanni Vigna},
	year         = 2021,
	month        = {May},
	journal      = {arXiv (Cornell University)},
	booktitle    = {Proceedings of the IEEE Symposium on Security and Privacy (IEEE S\&P)},
	publisher    = {Cornell University},
	pages        = {–},
	doi          = {10.48550/arxiv.2104.08638},
	url          = {https://arxiv.org/pdf/2104.08638},
	note         = {arXiv preprint arXiv:2104.08638, accessed \url{https://arxiv.org/abs/2104.08638}}
}

@inproceedings{mishra2025totalsol,
	title        = {TotalSol: A Multi-Layer Static Analysis Method for Vulnerability Detection in Ethereum Based Smart Contracts},
	author       = {Mishra, Deepak Kumar and Mehra, Pawan Singh},
	year         = 2025,
	month        = 1,
	booktitle    = {2025 International Conference on Cognitive Computing in Engineering, Communications, Sciences and Biomedical Health Informatics (IC3ECSBHI)},
	publisher    = {IEEE},
	volume       = {},
	number       = {},
	pages        = {945--950},
	doi          = {10.1109/IC3ECSBHI63591.2025.10990789},
	url          = {https://doi.org/10.1109/ic3ecsbhi63591.2025.10990789}
}

@article{chen2022defectchecker,
	title        = {DefectChecker: Automated Smart Contract Defect Detection by Analyzing EVM Bytecode},
	author       = {Chen, Jiachi and Xia, Xin and Lo, David and Grundy, John and Luo, Xiapu and Chen, Ting},
	year         = 2022,
	month        = 7,
	journal      = {IEEE Transactions on Software Engineering},
	publisher    = {Institute of Electrical and Electronics Engineers (IEEE)},
	volume       = 48,
	number       = 7,
	pages        = {2189--2207},
	doi          = {10.1109/TSE.2021.3054928},
	issn         = {0098-5589}
}

@article{wang2019npchecker,
	title        = {Detecting nondeterministic payment bugs in Ethereum smart contracts},
	author       = {Wang, Shuai and Zhang, Chengyu and Su, Zhendong},
	year         = 2019,
	month        = {October},
	journal      = {Proc. ACM Program. Lang.},
	publisher    = {Association for Computing Machinery},
	address      = {New York, NY, USA},
	volume       = 3,
	number       = {OOPSLA},
	pages        = {1--29},
	doi          = {10.1145/3360615},
	issn         = {2475-1421},
	url          = {https://doi.org/10.1145/3360615},
	articleno    = 189,
	issue_date   = {October 2019},
	numpages     = 29
}

@misc{cccGithub,
	title        = {{G}it{H}ub - Fraunhofer-AISEC/cpg-contract-checker at \#4ff0493 --- github.com},
	author       = {},
	year         = {},
	note         = {Accessed November 2025},
	howpublished = {\url{https://github.com/Fraunhofer-AISEC/cpg-contract-checker}}
}

@misc{confuzziusGithub,
	title        = {{G}it{H}ub - christoftorres/ConFuzzius at v0.0.2 --- github.com},
	author       = {},
	year         = {},
	note         = {Accessed November 2025},
	howpublished = {\url{https://github.com/christoftorres/ConFuzzius}}
}

@inproceedings{veloso2023conkas,
	title        = {Conkas: A Modular and Static Analysis Tool for Ethereum Smart Contracts},
	author       = {Nuno Veloso},
	year         = 2023,
	booktitle    = {Master’s Thesis, Instituto Superior Tecnico, Universidade de Lisboa},
	note         = {Available as technical report via \url{https://fenix.tecnico.ulisboa.pt/downloadFile/1689244997262417/94080-Nuno-Veloso_resumo.pdf}}
}

@misc{manticoreGithub,
	title        = {{G}it{H}ub - trailofbits/manticore at v0.3.7 --- github.com},
	author       = {},
	year         = {},
	note         = {Accessed November 2025},
	howpublished = {\url{https://github.com/trailofbits/manticore}}
}

@misc{mossberg2019manticore,
	title        = {Manticore: A User-Friendly Symbolic Execution Framework for Binaries and Smart Contracts},
	author       = {Mark Mossberg and Felipe Manzano and Eric Hennenfent and Alex Groce and Gustavo Grieco and Josselin Feist and Trent Brunson and Artem Dinaburg},
	year         = 2019,
	month        = 11,
	booktitle    = {2019 34th IEEE/ACM International Conference on Automated Software Engineering (ASE)},
	publisher    = {IEEE},
	pages        = {1186--1189},
	doi          = {10.1109/ase.2019.00133},
	url          = {https://arxiv.org/abs/1907.03890},
	archiveprefix = {arXiv},
	eprint       = {1907.03890},
	primaryclass = {cs.SE}
}

@misc{oyentePlusGithub,
	title        = {{G}it{H}ub - smartbugs/oyente-plus at v1.0.0 --- github.com},
	author       = {},
	year         = {},
	note         = {Accessed November 2025},
	howpublished = {\url{https://github.com/smartbugs/oyente_plus}}
}

@misc{securifyGithub,
	title        = {{G}it{H}ub - eth-sri/securify at \#51ba124 --- github.com},
	author       = {},
	year         = {},
	note         = {Accessed November 2025},
	howpublished = {\url{https://github.com/eth-sri/securify}}
}

@misc{securify2Github,
	title        = {{G}it{H}ub - eth-sri/securify2 at \#1913dfe --- github.com},
	author       = {},
	year         = {},
	note         = {Accessed November 2025},
	howpublished = {\url{https://github.com/eth‑sri/securify2}}
}

@misc{sfuzzGithub,
	title        = {{G}it{H}ub - duytai/sFuzz at \#ce87440 --- github.com},
	author       = {},
	year         = {},
	note         = {Accessed November 2025},
	howpublished = {\url{https://github.com/duytai/sFuzz}}
}

@misc{slitherGithub,
	title        = {{G}it{H}ub - crytic/slither at v0.11.3 --- github.com},
	author       = {},
	year         = {},
	note         = {Accessed November 2025},
	howpublished = {\url{https://github.com/crytic/slither}}
}

@inproceedings{tikhomirov2018smartcheck,
	title        = {SmartCheck: Static Analysis of Solidity Smart Contracts},
	author       = {Tikhomirov, Sergey and Tikhomirov, Sergey and Frolov, Alexey},
	year         = 2018,
	booktitle    = {arXiv preprint arXiv:1807.03325}
}

@misc{solhintGithub,
	title        = {{G}it{H}ub - protofire/solhint at v6.0.0 --- github.com},
	author       = {},
	year         = {},
	note         = {Accessed November 2025},
	howpublished = {\url{https://github.com/protofire/solhint}}
}

@misc{teetherGithub,
	title        = {{G}it{H}ub - nescio007/teether at \#04adf56 --- github.com},
	author       = {},
	year         = {},
	note         = {Accessed November 2025},
	howpublished = {\url{https://github.com/nescio007/teether}}
}

@inproceedings{krupp2018teether,
	title        = {teEther: Gnawing at Ethereum to Automatically Exploit Smart Contracts},
	author       = {Johannes Krupp and Christian Rossow},
	year         = 2018,
	month        = {August},
	journal      = {USENIX Security Symposium},
	booktitle    = {27th USENIX Security Symposium (USENIX Security ’18)},
	publisher    = {USENIX Association},
	address      = {Baltimore, MD, USA},
	pages        = {1317--1333},
	isbn         = {978‑1‑939133‑04‑5},
	url          = {https://www.usenix.org/conference/usenixsecurity18/presentation/krupp},
	editor       = {Enck, William and Felt, Adrienne Porter}
}

@misc{vandalGithub,
	title        = {{G}it{H}ub - usyd-blockchain/vandal at \#d2b0043 --- github.com},
	author       = {},
	year         = {},
	note         = {Accessed November 2025},
	howpublished = {\url{https://github.com/usyd-blockchain/vandal}}
}

@article{brent2018vandal,
	title        = {Vandal: A Scalable Security Analysis Framework for Smart Contracts},
	author       = {Lexi Brent and Anton Jurisevic and Michael Kong and Eric Liu and Fran\c{c}ois Gauthier and Vincent Gramoli and Ralph Holz and Bernhard Scholz},
	year         = 2018,
	month        = 9,
	journal      = {arXiv preprint arXiv:1809.03981},
	publisher    = {Cornell University},
	doi          = {10.48550/arxiv.1809.03981},
	url          = {https://arxiv.org/abs/1809.03981},
	note         = {Preprint; see project page for code and artifacts}
}

@article{weiss2024analyzing,
	title        = {Analyzing the Impact of Copying-and-Pasting Vulnerable Solidity Code Snippets from Question-and-Answer Websites},
	author       = {Konrad Weiss and Christof Ferreira Torres and Florian Wendland},
	year         = 2024,
	month        = 11,
	journal      = {arXiv preprint arXiv:2409.07586},
	booktitle    = {Proceedings of the 2024 ACM on Internet Measurement Conference, IMC 2024, Madrid, Spain, November 4-6, 2024},
	publisher    = {ACM},
	pages        = {713--730},
	url          = {https://arxiv.org/abs/2409.07586},
	note         = {Preprint; contains links to tools CCC and CCD},
	editor       = {Vallina-Rodriguez, Narseo and Suarez-Tangil, Guillermo and Levin, Dave and Pelsser, Cristel}
}

@inproceedings{slither,
	title        = {Slither: A static analysis framework for smart contracts},
	author       = {Feist, Josselin and Grieco, Gustavo and Groz, Bogdan},
	year         = 2019,
	month        = 5,
	booktitle    = {Proceedings of the 2nd International Workshop on Emerging Trends in Software Engineering for Blockchain (WETSEB)},
	publisher    = {IEEE},
	pages        = {8--15},
	doi          = {10.1109/wetseb.2019.00008},
	url          = {https://arxiv.org/pdf/1908.09878}
}

@inproceedings{oyente,
	title        = {Making smart contracts smarter},
	author       = {Luu, Loi and Chu, Duc-Hiep and Olickel, Hrishi and Saxena, Prateek and Hobor, Aquinas},
	year         = 2016,
	month        = 10,
	journal      = {Proceedings of the 2022 ACM SIGSAC Conference on Computer and Communications Security},
	booktitle    = {Proceedings of the ACM SIGSAC Conference on Computer and Communications Security (CCS)},
	publisher    = {ACM},
	pages        = {254--269},
	editor       = {Weippl, Edgar R. and Katzenbeisser, Stefan and Kruegel, Christopher and Myers, Andrew C. and Halevi, Shai}
}

@misc{huanggai,
	title        = {{G}it{H}ub - xf97/{H}uang{G}ai at v1.0.0 --- github.com},
	author       = {},
	year         = {},
	note         = {[Accessed 29-05-2025]},
	howpublished = {\url{https://github.com/xf97/HuangGai/tree/v1.0.0}}
}

@misc{openzeppelinOpenZeppelinDocs,
	author = {},
	title = {{E}{R}{C} 20 - {O}pen{Z}eppelin {D}ocs --- docs.openzeppelin.com},
	howpublished = {\url{https://docs.openzeppelin.com/contracts/2.x/api/token/erc20}},
	year = {},
	note = {[Accessed 30-05-2025]},
}

@misc{githubGitHubSmartbugsconkas,
	author = {},
	title = {{G}it{H}ub - smartbugs/conkas: {E}thereum {V}irtual {M}achine ({E}{V}{M}) {B}ytecode or {S}olidity {S}mart {C}ontract static analysis tool based on symbolic execution --- github.com},
	howpublished = {\url{https://github.com/smartbugs/conkas}},
	year = {},
	note = {[Accessed 30-05-2025]},
}

@misc{githubGitHubConsenSysDiligencemythril,
	author = {},
	title = {{G}it{H}ub - {C}onsen{S}ys{D}iligence/mythril: {M}ythril is a symbolic-execution-based securty analysis tool for {E}{V}{M} bytecode. {I}t detects security vulnerabilities in smart contracts built for {E}thereum and other {E}{V}{M}-compatible blockchains. --- github.com},
	howpublished = {\url{https://github.com/ConsenSysDiligence/mythril}},
	year = {},
	note = {[Accessed 30-05-2025]},
}

@article{chen2025chatgpt,
  title={When chatgpt meets smart contract vulnerability detection: How far are we?},
  author={Chen, Chong and Su, Jianzhong and Chen, Jiachi and Wang, Yanlin and Bi, Tingting and Yu, Jianxing and Wang, Yanli and Lin, Xingwei and Chen, Ting and Zheng, Zibin},
  journal={ACM Transactions on Software Engineering and Methodology},
  volume={34},
  number={4},
  pages={1--30},
  year={2025},
  publisher={ACM New York, NY}
}

@article{ding2025smartguard,
  title={SmartGuard: An LLM-enhanced framework for smart contract vulnerability detection},
  author={Ding, Hao and Liu, Yizhou and Piao, Xuefeng and Song, Huihui and Ji, Zhenzhou},
  journal={Expert Systems with Applications},
  volume={269},
  pages={126479},
  year={2025},
  publisher={Elsevier}
}

@inproceedings{choi2021smartian,
  title={Smartian: Enhancing smart contract fuzzing with static and dynamic data-flow analyses},
  author={Choi, Jaeseung and Kim, Doyeon and Kim, Soomin and Grieco, Gustavo and Groce, Alex and Cha, Sang Kil},
  booktitle={2021 36th IEEE/ACM International Conference on Automated Software Engineering (ASE)},
  pages={227--239},
  year={2021},
  organization={IEEE}
}

@article{wei2025advanced,
  title={Advanced smart contract vulnerability detection via llm-powered multi-agent systems},
  author={Wei, Zhiyuan and Sun, Jing and Sun, Yuqiang and Liu, Ye and Wu, Daoyuan and Zhang, Zijian and Zhang, Xianhao and Li, Meng and Liu, Yang and Li, Chunmiao and others},
  journal={IEEE Transactions on Software Engineering},
  year={2025},
  publisher={IEEE}
}

@inproceedings{sun2024gptscan,
  title={Gptscan: Detecting logic vulnerabilities in smart contracts by combining gpt with program analysis},
  author={Sun, Yuqiang and Wu, Daoyuan and Xue, Yue and Liu, Han and Wang, Haijun and Xu, Zhengzi and Xie, Xiaofei and Liu, Yang},
  booktitle={Proceedings of the IEEE/ACM 46th International Conference on Software Engineering},
  pages={1--13},
  year={2024}
}

@article{boi2024smart,
  title={Smart contract vulnerability detection: The role of large language model (llm)},
  author={Boi, Biagio and Esposito, Christian and Lee, Sokjoon},
  journal={ACM SIGAPP applied computing review},
  volume={24},
  number={2},
  pages={19--29},
  year={2024},
  publisher={ACM New York, NY, USA}
}

@inproceedings{wu2024advscanner,
  title={Advscanner: Generating adversarial smart contracts to exploit reentrancy vulnerabilities using llm and static analysis},
  author={Wu, Yin and Xie, Xiaofei and Peng, Chenyang and Liu, Dijun and Wu, Hao and Fan, Ming and Liu, Ting and Wang, Haijun},
  booktitle={Proceedings of the 39th IEEE/ACM International Conference on Automated Software Engineering},
  pages={1019--1031},
  year={2024}
}

@article{ouyang2025empirical,
  title={An empirical study of the non-determinism of chatgpt in code generation},
  author={Ouyang, Shuyin and Zhang, Jie M and Harman, Mark and Wang, Meng},
  journal={ACM Transactions on Software Engineering and Methodology},
  volume={34},
  number={2},
  pages={1--28},
  year={2025},
  publisher={ACM New York, NY}
}

@article{feichtinger2024sok,
  title={Sok: Attacks on daos},
  author={Feichtinger, Rainer and Fritsch, Robin and Heimbach, Lioba and Vonlanthen, Yann and Wattenhofer, Roger},
  journal={arXiv preprint arXiv:2406.15071},
  year={2024}
}

@inproceedings{atzei2017survey,
  title={A Survey of Attacks on Ethereum Smart Contracts ({SoK})},
  author={Atzei, Nicola and Bartoletti, Massimo and Cimoli, Tiziana},
  booktitle={Principles of Security and Trust (POST)},
  series={LNCS},
  volume={10204},
  pages={164--186},
  year={2017},
  publisher={Springer},
  doi={10.1007/978-3-662-54455-6_8}
}

@inproceedings{kalra2018zeus,
  title={Zeus: analyzing safety of smart contracts.},
  author={Kalra, Sukrit and Goel, Seep and Dhawan, Mohan and Sharma, Subodh},
  booktitle={Ndss},
  pages={1--12},
  year={2018}
}

@inproceedings{zhang2020txspector,
  author    = {Mengya Zhang and Xiaokuan Zhang and Yinqian Zhang and Zhiqiang Lin},
  title     = {{TXSPECTOR}: Uncovering Attacks in Ethereum from Transactions},
  booktitle = {29th USENIX Security Symposium (USENIX Security 20)},
  year      = {2020},
  pages     = {2775--2792},
  publisher = {USENIX Association},
  month     = aug,
  isbn      = {978-1-939133-17-5},
  url       = {https://www.usenix.org/conference/usenixsecurity20/presentation/zhang-mengya}
}

@inproceedings{ferreira2020aegis,
  title={{\AE}gis: Shielding vulnerable smart contracts against attacks},
  author={Ferreira Torres, Christof and Baden, Mathis and Norvill, Robert and Fiz Pontiveros, Beltran Borja and Jonker, Hugo and Mauw, Sjouke},
  booktitle={Proceedings of the 15th ACM Asia Conference on Computer and Communications Security},
  pages={584--597},
  year={2020}
}

@article{salzano2025empirical,
  title={An Empirical Analysis of Vulnerability Detection Tools for Solidity Smart Contracts Using Line Level Manually Annotated Vulnerabilities},
  author={Salzano, Francesco and Antenucci, Cosmo Kevin and Scalabrino, Simone and Rosa, Giovanni and Oliveto, Rocco and Pareschi, Remo},
  journal={arXiv preprint arXiv:2505.15756},
  year={2025}
}

@inproceedings{liu2025reentrancy,
  title={Reentrancy Redux: The Evolution of Real-World Reentrancy Attacks on Blockchains},
  author={Liu, Yuqi and Xi, Rui and Pattabiraman, Karthik},
  booktitle={2025 55th Annual IEEE/IFIP International Conference on Dependable Systems and Networks (DSN)},
  pages={576--588},
  year={2025},
  organization={IEEE}
}

\end{document}